\documentclass[aps,prd,superscriptaddress,twoside,twocolumn,nofootinbib,showpacs,
floatfix]{revtex4-2}
\usepackage{amsmath,amssymb}
\usepackage{graphicx,bm}
\usepackage{slashed,color}
\usepackage[colorlinks=true,pdfstartview=FitV,bookmarks=true,bookmarksnumbered=true,
breaklinks]{hyperref}

\hypersetup{linkcolor=red, citecolor=blue, urlcolor=blue}

\usepackage[normalem]{ulem} % \sout{old text} for strikeout
\usepackage[dvipsnames]{xcolor} % For blue in-text comments and
\begin{document}
%-------------------------------------------------------------------------------------
\title{\boldmath Effective Lagrangian for strong and electromagnetic interactions of
high-spin resonances}
%-------------------------------------------------------------------------------------
\author{Sang-Ho Kim}
\email[E-mail: ]{shkimphy@gmail.com}
\affiliation{Department of Physics and Origin of Matter and Evolution of Galaxies
(OMEG) Institute, Soongsil University, Seoul 06978, Korea}
%-------------------------------------------------------------------------------------
\author{Yongseok Oh}
\email[ ]{Deceased April 3, 2023}
%\affiliation{ }
\affiliation{Department of Physics, Kyungpook National University, Daegu 41566, Korea}
%-------------------------------------------------------------------------------------
\author{Sangyeong Son}
\affiliation{Department of Physics, Kyungpook National University, Daegu 41566, Korea}
%-------------------------------------------------------------------------------------
\author{S. Sakinah}
\affiliation{Department of Physics, Kyungpook National University, Daegu 41566, Korea}
%-------------------------------------------------------------------------------------
\author{Myung-Ki Cheoun}
\affiliation{Department of Physics and Origin of Matter and Evolution of Galaxies
(OMEG) Institute, Soongsil University, Seoul 06978, Korea}
%-------------------------------------------------------------------------------------

\date{\today}

\begin{abstract}
Recent experiments of photon-nucleon and meson-nucleon scatterings have accumulated
a lot of data for various meson production processes.
One of the purposes of those experiments is to search for the missing resonances
which are not discovered until now but whose existence was predicted by hadron
models.
The analyses of the data requires the development of dynamical coupled-channel
models.
Since several missing resonances are expected to have spin higher than $3/2$, we need
to include higher-spin resonances in dynamical coupled-channel models, which enable
us to determine the couplings of effective Lagrangians of higher-spin baryons with
pseudoscalar mesons or vector mesons. 
However, hadron models, such as quark models, give predictions only of the decay
amplitudes of such baryons.
Here we demonstrate the formalism of high-spin resonances and construct the relation
between the coupling constants of effective Lagrangians and the partial decay
widths that can be predicted by hadron models. 
This allows us to compare the coupling constants to the hadron model predictions not
only in magnitude but in sign as well.
\end{abstract}

\maketitle
%-------------------------------------------------------------------------------------
\section{Introduction}
%-------------------------------------------------------------------------------------
Until today there have been extensive studies on meson production in the
photon-nucleon or meson-nucleon scattering at various electron/photon- and
meson-beam facilities. 
One of the aims of these investigations is to obtain a clue to solve the so-called
``missing resonance puzzle''.
In the review of Particle Data Group (PDG)~\cite{PDG:2024cfk}, there are about 27
nucleon resonances and 22 $\Delta$ resonances.
However, many phenomenological models on baryon structure such as quark potential
models predict much richer spectrum of such resonances~\cite{Capstick:2000qj}. 
Hence, many resonances are still missing and yet to be discovered.
This discrepancy has been attributed to the weak coupling of missing resonances to the
$\pi N$ channel, which has offered a way to discover most resonances reported in the
PDG.
Therefore, there have been continuos efforts to find missing resonances in other
reactions such as vector-meson photoproduction~\cite{CLAS:2013qgi,Denisenko:2016ugz,
CLAS:2017sgi,CLAS:2018drk,Wei:2019imo}, $\eta$-meson photoproduction~\cite{
A2:2014pie,A2:2014pie,A2:2016bij,CLAS:2020vpr,LEPS2BGOegg:2022dop,CBELSATAPS:2023ucy},
and $K$-meson photoproduction~\cite{CLAS:2006pde,CLAS:2009rdi,CLAS:2009fmu,
CLAS:2010aen,LEPS:2017pzl,A2:2018doh}.
Through these studies, one hopes to discover such missing resonances by analyzing the
scattering data in the resonance region.
Similar problems are also found for the spectra of strange hyperons and heavy-flavor
hadrons. 

In addition to the ``missing resonance puzzle'', another major research topic is to
constrain and understand the properties of resonances.
The properties of most resonances listed in the PDG are not well known including the
spin-parity quantum numbers and decay widths.
In some cases even the mass of a resonance could be estimated with a very wide range.
This is because of the model dependence of resonance parameters when extracted from
the experimental data.

Constraining the basic properties of resonances such as masses, spin-parity quantum
numbers, and total and partial decay widths is crucial to understand the strong and
electromagnetic (EM) interactions through the study of baryon structure because of
the strong model dependence of the resonance spectrum.
The success of most phenomenological models to describe the masses of ground state
octet and decuplet baryons is mainly based on the group structure of the flavor
symmetry and its breaking.
On the other hand, the predicted mass spectrum of excited states is strongly
dependent on the assumed dynamics and, therefore, is highly model dependent
(see, for example, Refs.~\cite{Capstick:2000qj,Oh:2007cr}).
Understanding the properties of known resonances together discovering new resonances
is, therefore, expected to shed light on the understanding of hadron structure.

Among the known resonances, most of the resonances with a low mass have spin-$1/2$
or $3/2$.
However, one can easily find higher-spin states for higher-mass resonances and even
the resonance with spin-$15/2$ is listed in the PDG.
Therefore, it is required to include high-spin resonances in the analyses of the
photon-nucleon and meson-nucleon scattering data.
Indeed, there have been efforts to include spin-5/2 baryons in the analyses of
scattering data~\cite{Shklyar:2004dy}.
Furthermore, the recent analyses of $\Xi$ production via photo-~\cite{Man:2011np}
and $K$-~\cite{Feijoo:2015yja,Jackson:2015dva,Kim:2023jij} beams show the important
role of a spin-7/2 hyperon resonance in the reaction mechanism.
All these works show the importance of including high-spin resonances in anaylzing
the experimental observables.
The method to include spin-$3/2$ baryons in the mechanism of meson-production
processes has been studied by many authors since the seminal work of Rarita and
Schwinger~\cite{Rarita:1941mf}.
However, the interactions and decays of resonances with a higher spin are not
well-known.
It is thus required to investigate the formalism for describing the interactions
and decays of high-spin resonances.
Such efforts can be found, for example, in Ref.~\cite{Riska:2000gd}, where the
interaction Lagrangian is constructed based on a quark model.
In this work, we describe the general and relativistic form of the interaction
Lagrangian and partial decay width of a baryon  with an arbitrary spin based on the
Rarita-Schwinger formalism~\cite{Rarita:1941mf}.

Another motivation of this work is how to relate the coupling constants of
effective Lagrangians of resonances to the predictions of baryon models.
By constructing effective Lagrangians, we have to introduce coupling constants
which should be determined by the experimental data.
However, what we can know from experimental data is the partial decay widths of
resonances.
This allows us to estimate the magnitude of coupling constants but their signs are
not determined.
The signs of coupling constants are related to the relative phase of the wave
functions of the participating particles, which can constrain phenomenological
models for hadron structure.
Furthermore, the problem of sign ambiguity becomes crucial as we include more and
more intermediate states in reaction mechanisms.
In fact, the signs of coupling constants as well as their magnitudes should be
extracted by analyzing observed physical quantities with sophisticated
coupled-channel models.
(see, for example, Ref.~\cite{Matsuyama:2006rp}.)
Even in this model, however, since there are many coupling constants, it is important
to know the  meaningful starting values to avoid local minima during the fitting
procedure.
Thus, if we believe that some phenomenological models can provide reasonable range
of coupling constants, we can use such predictions of the starting guess for
coupling constants in coupled-channel approaches.
In many phenomenological models for hadron structure, the wavefunctions of
resonances are identified, which leads to the predictions of the decay amplitudes.
Therefore, we need to know the relationship between the coupling constants of
effective Lagrangians and the predictions of hadron models on the decay amplitudes.
This will be done in this work and the explicit relation between the two will be made.

Stemming from the approach to construct the Rarita-Schwinger
spin-vector~\cite{Rarita:1941mf},
this field consists of the two irreducible representations for spin-3/2 and
spin-1/2 degrees of freedom.
Those are reduced to the physical spin degrees of freedom describing the on-shell
spin-3/2 particle with the help of the Euler-Lagrange field equations. 
However, for the spin-3/2 field off the mass shell, one encounters the
inconsistent coupling constants accompanying the unphysical spin-1/2 backgrounds
for interactions such as the $\Delta\to\pi N$ decay. 
Several approaches have been suggested to address the consistent interaction
involving the pure spin-3/2 field, with the spin-1/2 contributions decoupled
from observables (see, for a review, Sec.~IV of Ref.~\cite{Pascalutsa:2006up} and
references therein). 
Such efforts include the construction of interaction Lagrangians invariant under
the point transformation~\cite{Peccei:1968bct,Haberzettl:1998rw,Pilling:2004wk,
Pilling:2004cu,Wies:2006rv}, the gauge
transformation~\cite{Pascalutsa:1998pw,Pascalutsa:1999zz}, and the unification
thereof for high-spin fields~\cite{Vereshkov:2013kma}.
It was also shown that any inconsistent coupling can be transformed into consistent
coupling via a field redefinition and the equivalence
theorem~\cite{Pascalutsa:2000kd}.

This paper is organized as follows.
In Sec.~\ref{Sec:II}, we first review the propagators of arbitrary spin fields.
This is required to calculate the partial decay widths and to construct the reaction
amplitudes for scattering processes.
Here we give the general expression for the propagators for bosons as well as for
fermions.
It also contains explicit formulae for the propagators of several high-spin fields.
Then, in Sec.~\ref{Sec:III}, we construct effective Lagrangians of arbitrary spin
resonances which couple to a meson and a spin-1/2 baryon, i.e., the nucleon.
Here we consider their couplings to pseudoscalar mesons and to vector mesons.
Namely, we construct the general effective Lagrangians for $j^P \to 0^- + 1/2^+$ and
for $j^P \to 1^- + 1/2^+$.
It is then straightforward to extend the formulae to the cases of scalar and
pseudovector mesons.
The explicit formulae for the decay amplitudes of resonances are given in terms of
the coupling constants of effective Lagrangians.
Section~\ref{Sec:IV} applies the process of $j^P \to 1^- + 1/2^+$ to the
photo-decay of resonances into the nucleon.
Because of the absence of the helicity-0 state, the number of independent couplings
reduces by one.

Recent activities also show that the production of decuplet baryons would have
non-negligible cross sections compared to that of octet baryon
production~\cite{CLAS:2013rxx,CLAS:2018kvn}.
Therefore, it is necessary to have the formalism for the interactions and decays
of a spin-3/2 baryon.
Section~\ref{Sec:V} contains those information for $j^P \to 0^- + 3/2^+$ and for
$j^P \to 1^- + 3/2^+$.
Finally, we summarize in Sec.~\ref{sec:summary}.
Several explicit formulae for propagators of the fields are given in Appendix.

%-------------------------------------------------------------------------------------
\section{Propagators} \label{Sec:II}
%-------------------------------------------------------------------------------------

In this section, we review the general form of the propagators of a field with an
arbitrary spin.
The general expression for propagators can be found in
Refs.~\cite{Behrends:1957rup,Fronsdal58,Rushbrooke:1966zz,Chang:1967zzc}.
We here summarize the main results for the form of the propagators referring the
details to Refs.~\cite{Behrends:1957rup,Fronsdal58,Rushbrooke:1966zz,Chang:1967zzc}.

In the Rarita-Schwinger formalism, a boson field of spin-$j$ is described by a tensor
of rank $n=j$, $R_{\alpha_1^{} \alpha_2^{} \cdots \alpha_n^{}}$, which satisfies a free field
equation,
%EQUATION>>>
\begin{align}
(\partial_\mu \partial^\mu + M^2) R_{\alpha_1^{} \alpha_2^{} \cdots \alpha_n^{}} = 0,
\end{align}
%EQUATION<<<
where $M$ is its mass.
This field is symmetric under the exchange of two indices, i.e.,
$R_{\alpha_1^{} \cdots \alpha_i^{} \cdots \alpha_j^{} \cdots \alpha_n^{}}
= R_{\alpha_1^{} \cdots \alpha_j^{} \cdots \alpha_i^{} \cdots \alpha_n^{}}$,
and satisfies the subsidiary conditions
%EQUATION>>>
\begin{align}
& p^{\alpha_1^{}} R_{\alpha_1^{} \alpha_2^{} \cdots \alpha_n^{}} = 0,   \cr
& g^{\alpha_1^{}\alpha_2^{}} R_{\alpha_1^{} \alpha_2^{} \cdots \alpha_n^{}} = 0,
\label{eq:subsidiary1}
\end{align}
%EQUATION<<<
where $p^\alpha$ is the four-momentum of the field.

A fermion field of spin-$j$ is represented by a tensor of rank $n=j-1/2$, which
contains a Dirac spinor. Its free field equation is
%EQUATION>>>
\begin{align}
(i \slashed{\partial} - M) R_{\alpha_1^{} \alpha_2^{} \cdots \alpha_n^{}} = 0.
\end{align}
%EQUATION<<<
Again, this field is symmetric under the exchange of two indices.
In addition to the subsidiary conditions of Eq.~(\ref{eq:subsidiary1}), the fermion
field has an additional constraint as follows
%EQUATION>>>
\begin{align}
\gamma^{\alpha_1^{}} R_{\alpha_1^{} \alpha_2^{} \cdots \alpha_n^{}} = 0.
\end{align}
%EQUATION<<<

The general form of the propagator is
%EQUATION>>>
\begin{align}
S(p) = \frac{1}{p^2 - M^2} \Delta_{\alpha_1^{} \cdots \alpha_n^{}}^{\beta_1^{} \cdots \beta_n^{}}
\end{align}
%EQUATION<<<
for a boson field and
%EQUATION>>>
\begin{align}
S(p) = \frac{1}{p^2 - M^2} (\slashed{p} + M) 
\Delta_{\alpha_1^{} \cdots \alpha_n^{}}^{\beta_1^{} \cdots \beta_n^{}}
\end{align}
%EQUATION<<<
for a fermion field.
The projection operator $\Delta$ has upper and lower indices through its definition,
which reads
%EQUATION>>>
\begin{align}
\sum_{\rm spin} R_{\alpha_1^{} \cdots \alpha_n^{}} R^{\beta_1^{} \cdots \beta_n^{}} = 
\Lambda_\pm \Delta_{\alpha_1^{} \cdots \alpha_n^{}}^{\beta_1^{} \cdots \beta_n^{}},
\end{align}
%EQUATION<<<
for both the boson and fermion fields, where $\Lambda_\pm = (M \pm \slashed{p})/2M$ 
are the usual energy projection operators of the spin-1/2 Dirac field in the case
of a fermion and $\Lambda_\pm = 1$ in the case of a boson.

In Refs.~\cite{Behrends:1957rup,Fronsdal58,Rushbrooke:1966zz,Chang:1967zzc}, the
general expression for the projection operator for a spin-$j$ field with a momentum
$p$, i.e., $\Delta_{\alpha_1^{} \cdots \alpha_n^{}}^{\beta_1^{} \cdots \beta_n^{}}(j,p)$, is
dervied.
For a boson of spin $j=n$, we have
%------------------------------>>> widetext
\begin{widetext}
%EQUATION>>>
\begin{align}
\Delta_{\alpha_1^{} \cdots \alpha_n^{}}^{\beta_1^{} \cdots \beta_n^{}}(j,p) = 
\left( \frac{1}{n!} \right)^2 \sum_{P(\alpha), P(\beta)}
\left[ \prod_{i=1}^n \bar{g}_{\alpha_i^{}}^{\beta_i^{}} 
+ a_1^{(n)} \bar{g}_{\alpha_1^{} \alpha_2^{}} \bar{g}^{\beta_1^{} \beta_2^{}}
\prod_{i=3}^n \bar{g}_{\alpha_i^{}}^{\beta_i^{}} 
+ \cdots
+ a_{n/2}^{(n)} \bar{g}_{\alpha_1^{} \alpha_2^{}} \bar{g}^{\beta_1^{} \beta_2^{}}
\cdots \bar{g}_{\alpha_{n-1}^{} \alpha_n^{}} \bar{g}^{\beta_{n-1}^{} \beta_n^{}}
\right],
\label{eq:prop-even}
\end{align}
%EQUATION<<<
if $n=j$ is an even integer, and
%EQUATION>>>
\begin{align}
\Delta_{\alpha_1^{} \cdots \alpha_n^{}}^{\beta_1^{} \cdots \beta_n^{}}(j,p) &= 
- \left( \frac{1}{n!} \right)^2 \sum_{P(\alpha), P(\beta)}
\left[ \prod_{i=1}^n \bar{g}_{\alpha_i^{}}^{\beta_i^{}} 
+ a_1^{(n)} \bar{g}_{\alpha_1^{} \alpha_2^{}} \bar{g}^{\beta_1^{} \beta_2^{}}
\prod_{i=3}^n \bar{g}_{\alpha_i^{}}^{\beta_i^{}} 
+ \cdots \right.   \cr
& \left. \qquad \qquad \qquad \qquad \quad \mbox{}
+ a_{(n-1)/2}^{(n)} \bar{g}_{\alpha_1^{} \alpha_2^{}} \bar{g}^{\beta_1^{} \beta_2^{}}
\cdots \bar{g}_{\alpha_{n-2}^{} \alpha_{n-1}^{}} \bar{g}^{\beta_{n-2}^{} \beta_{n-1}^{}}
\bar{g}_{\alpha_n^{}}^{\beta_n^{}}
\right],
\label{eq:prop-odd}
\end{align}
%EQUATION<<<
\end{widetext}
%-------------------------------<<< widetext
if $n=j$ is an odd integer\footnote{References~\cite{Behrends:1957rup,Fronsdal58,
Rushbrooke:1966zz,Chang:1967zzc} use the convention of $g^{\mu\nu} =
\mathrm{diag}(-1,1,1,1)$ while we adopt $g^{\mu\nu} = \mathrm{diag}(1,-1,-1,-1)$.},
where the sum is over all possible permutations of $\alpha_i^{}$'s and $\beta_i^{}$'s.
For a fermion of spin $j = n-1/2$, it is written as
%EQUATION>>>
\begin{align}
\Delta_{\alpha_1^{} \cdots \alpha_{n-1}^{}}^{\beta_1^{} \cdots \beta_{n-1}^{}}(j,p) =
- \frac{n}{2n+1} \gamma^\alpha \gamma_\beta^{} 
\Delta_{\alpha \alpha_1^{} \cdots \alpha_{n-1}^{}}^{\beta \beta_1^{} \cdots \beta_{n-1}^{}}
(j+\textstyle\frac12,p).
\label{eq:prop-fermion}
\end{align}
%EQUATION<<<
Here, the constant $a_r^{(n)}$ is defined as
%EQUATION>>>
\begin{align}
a_r^{(n)} =& \left( -\frac12 \right)^r \frac{n!}{r!(n-2r)!}   \cr
& \times \frac{1}{(2n-1)(2n-3) \cdots (2n-2r+1)}.
\end{align}
%EQUATION<<<
In Eqs.~(\ref{eq:prop-even})-(\ref{eq:prop-odd}), $\bar{g}_{\mu\nu}$ is defined as
%EQUATION>>>
\begin{align}
\bar{g}_{\mu\nu} = g_{\mu\nu} - \frac{1}{{p^2}} p_\mu^{} p_\nu^{}.
\end{align}
%EQUATION<<<
It is also convenient to define
%EQUATION>>>
\begin{align}
\bar{\gamma}^\mu = \gamma^\nu \bar{g}_\nu^\mu = \gamma^\mu - \frac{1}{{p^2}} \slashed{p}
p^\mu.
\end{align}
%EQUATION<<<
It is then straightforward to obtain the explicit expression for 
$\Delta_{\alpha_1^{} \cdots \alpha_n^{}}^{\beta_1^{} \cdots \beta_n^{}}(j,p)$ of a given spin
and those for several cases are given in Appendix.

%-------------------------------------------------------------------------------------
\section{Couplings of baryon resonances into a meson and a
spin-$1/2$ baryon} \label{Sec:III}
%-------------------------------------------------------------------------------------

In this section, we construct the general form of the $RN\pi$ and $RNV$ interactions, 
where $R$ stands for a baryon resonance of spin-parity $j^P$ and $V$ denotes a vector
meson.
The effective Lagrangians obtained for these interactions describe the decays of
$j^P \to 0^- + 1/2^+$ and $j^P \to 1^- + 1/2^+$.
Throughout this paper, we do \textit{not} consider the isospin factor.
The isospin factor depends on the flavor structure as shown, for example, in
Ref.~\cite{Oh:2004gz}.
Since the final particle has $j^P = 1/2^+$, we represent it by the nucleon.

\subsection{Pseudoscalar meson couplings}

\begin{figure}[h]
\centering
\includegraphics[width=0.25\textwidth,angle=0,clip]{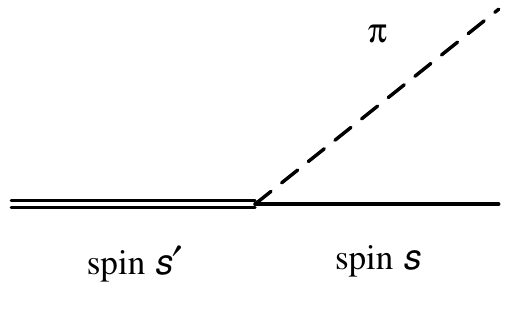}
\caption{Vertex of the pion decay of a spin-$s'$ baryon into a spin-$s$ baryon.}
\label{FIG01}
\end{figure}

The number of independent coupling constants (or form factors) of the decay of a
spin-$s'$ particle into pion and a spin-$s$ particle, depicted in Fig.~\ref{FIG01},
can be obtained by considering angular momentum conservation and invariance under
$P$ and $T$ transformations~\cite{Durand:1962zza}. 
Shown in Table~\ref{TAB01} are the results for the two cases~\cite{Durand:1962zza},
namely, when the particles other than the pion are fermions and bosons.
It is then clear that, for the decay of $j^P \to 0^- + 1/2^+$, there is only one
form factor as $j_{\rm min}^{} = \min(s,s')$ is always $1/2$.

%FIGURE>>>
%FIGURE<<<

%TABLE>>>
\begin{table}[h]
\centering
\begin{tabular}{c|cc} \hline\hline
 & $(-1)^\gamma = +1$ & $(-1)^\gamma = -1$ \\ \hline
Fermions & $j_{\rm min}^{} + \frac12$ & $j_{\rm min}^{} + \frac12$ \\
Bosons & $j_{\rm min}^{} + 1$ & $j_{\rm min}^{}$ \\ \hline\hline
\end{tabular}
\caption{Number of independent couplings for the pion decay of a spin-$s'$ particle
into a spin-$s$ particle~\cite{Durand:1962zza}.
The upper (lower) line is for a fermion (boson) decay.
Here $\gamma = s+s'+P+1$, where $(-1)^P = P_i P_f$, i.e., the relative parity of the
initial and final states whose spins are $s'$ and $s$, respectively, and $j_{\rm min}
= \min(s,s')$.}
\label{TAB01}
\end{table}
%TABLE<<<

The decay amplitude for $R \to N \pi$ can be written as~\cite{Kim:2018qfu,Suh:2018yiu}
%EQUATION>>>
\begin{align}
&\left < \pi({\bf k_\pi})\,N(-{\bf k_\pi},m_f) | -i \mathcal{H}_{\mathrm{int}} |
R({\bf 0},m_j) \right >
\cr   \hspace{-1em} =&
4 \pi M_R \sqrt\frac{2}{k_\pi} \sum_{\ell,m_\ell}
\langle \ell\, m_\ell\, {\textstyle\frac{1}{2}}\, m_f | j \,m_j \rangle Y_{\ell,m_\ell} 
({\bf \hat {k_\pi}}) G(\ell),
\label{eq:DA}
\end{align}
%EQUAITON<<<
where $\langle\ell\,m_\ell\,\frac{1}{2}\,m_f|j \,m_j \rangle$ and $Y_{\ell,m_\ell}
({\bf \hat {k_\pi}})$ are the Clebsch-Gordan coefficient and spherical harmonics,
respectively.
$k_\pi = |{\bf k}_\pi|$ is the magnitude of the three-momentum of the pion in the rest
frame of the resonance,
%EQUATION>>>
\begin{align}
k_\pi = \frac{1}{2M_R}
\sqrt{[M_R^2 - (M_N+M_\pi)^2][M_R^2 - (M_N-M_\pi)^2]},
\end{align}
%EQUATION<<<
where $M_h$ is the mass of a hadron $h$.

The partial-wave decay amplitude $G(\ell)$ is related to the decay width $\Gamma
(R \to N \pi)$ as follows
%EQUATION>>>
\begin{align}
\Gamma (R \to N \pi) = \sum_\ell |G(\ell)|^2 .
\label{eq:DA_Sum}
\end{align}
%EQUATION<<<
The spin-parity of the resonance places constraints on the relative orbital angular 
momentum $\ell$ of the $N \pi$ final state.
When $R$ has $j^P = \frac{1}{2}^-$, the relative orbital angular momentum is
restricted by the angular momentum conservation, so $s$ wave ($\ell = 0$) is only
allowed.
Similarly, when $R$ has $j^P = (1/2^+,\,3/2^+)$, $(3/2^-,\,5/2^-)$, $(5/2^+,7/2^+)$,
$(7/2^-,9/2^-)$, and $9/2^+$, the final state is in the relative $p$, $d$, $f$, $g$,
and $h$ waves, respectively.

We present the explicit form of the interaction Lagrangians and the relevant decay
widths and partial-wave decay amplitudes below.
For later use, we define
%EQUATION>>>
\begin{align}
\Gamma^{(\pm)} = \left( \begin{array}{c} \gamma_5^{} \\ 1 \end{array}
\right), \,
\Gamma^{(\pm)}_\mu = \left( \begin{array}{c} \gamma_\mu^{} \gamma_5^{} \\
\gamma_\mu^{} \end{array} \right).
\end{align}
%EQUATION<<<

\subsubsection{Spin-$1/2$ resonance}
When $R$ has $j^P = 1/2^\pm$, although there is only one form factor in this case,
one can adopt the pseudoscalar or pseudovector coupling for the interaction
Lagrangian as~\cite{Benmerrouche:1994uc}
%EQUATION>>>
\begin{align}
\mathcal{L}^{(1/2\pm)}_{RN\pi} =&
- g_{RN\pi}^{} \bar{N} \left[ i \lambda \Gamma^{(\pm)} 
\pi \mp \frac{1-\lambda}{M_R \pm M_N} \Gamma_\mu^{(\pm)}
\partial^\mu \pi \right] R
\cr   & 
\mp g_{RN\pi}^{} \bar{R} \left[ i \lambda \Gamma^{(\pm)} \pi -
\frac{1-\lambda}{M_R \pm M_N} \Gamma_\mu^{(\pm)} \partial^\mu \pi \right] N,
\label{lag:pi1}
\end{align}
%EQUATION<<<
where the upper sign is for a positive parity resonance and the lower for a negative
parity resonance. 
The parameter $\lambda$ is introduced for taking either the pseudovector ($\lambda=0$)
or pseudoscalar ($\lambda=1$) coupling as in the case of the $NN\pi$ coupling.
Since we are considering a resonance decay, both the initial and final baryons are on
their mass shell and the two coupling schemes are equivalent.

The obtained decay width for $R \to N\pi$ reads
\footnote{Note that we do not include the isospin factor.}
%EQUATION>>>
\begin{align}
\Gamma({\textstyle\frac12}^\pm \to N\pi) = \frac{g_{RN\pi}^2}{4\pi} \frac{k_\pi}{M_R}
(E_N \mp M_N),
\end{align}
%EQUATION<<<
where $E_N$ is the energy of the nucleon
%EQUATION>>>
\begin{align}
E_N = \sqrt{M_N^2 + k_\pi^2} = \frac{1}{2M_R^{}} \left( M_R^2 + M_N^2 - M_\pi^2 \right).
\end{align}
%EQUATION<<<

The partial-wave decay amplitude $G(\ell)$ derived from Eq.~(\ref{eq:DA}) is given by
%EQUATION>>>
\begin{align}
G\left( 1 \right) = -
\frac{1}{2\sqrt{\pi}} \sqrt{\frac{k_\pi}{M_R}} \sqrt{E_N - M_N}  g_{RN\pi},
\end{align}
%EQUATION<<<
for $j^P=1/2^+$ resonance, and
%EQUATION>>>
\begin{align}
G\left( 0 \right) =
\frac{1}{2\sqrt{\pi}} \sqrt{\frac{k_\pi}{M_R}} \sqrt{E_N + M_N}  g_{RN\pi},
\end{align}
%EQUATION<<<
for $j^P=1/2^-$ resonance.

\subsubsection{Spin-$3/2$ resonance}

The $RN\pi$ interaction where $R$ has $j^P = 3/2^\pm$ can be found, for example, in 
Ref.~\cite{Benmerrouche:1994uc}.
Since we are considering the decays of resonances, our results are independent of the
terms that contain the off-shell parameters.
Therefore, throughout this paper, we neglect the off-shell parameters and the
interaction Lagrangian reads
%EQUATION>>>
\begin{align}
\mathcal{L}^{(3/2\pm)}_{RN\pi} = 
\frac{g_{RN\pi}^{}}{M_\pi^{}}\, \bar{N}\, \Gamma^{(\mp)} \partial_\mu \pi R^\mu \pm
\frac{g_{RN\pi}^{}}{M_\pi^{}}\, \bar{R}^\mu\, \Gamma^{(\mp)} \partial_\mu \pi N.
\label{lag:pi3}
\end{align}
%EQUATION>>>
Then the decay width is obtained as
%EQUATION>>>
\begin{align}
\Gamma({\textstyle\frac32}^\pm \to N\pi) = 
\frac{1}{3} \frac{g_{RN\pi}^2}{4\pi} \frac{k_\pi^3}{M_R M_\pi^2} (E_N \pm M_N).
\end{align}
%EQUATION>>>

The obtained partial-wave decay amplitude is given by
%EQUATION>>>
\begin{align}
G\left( 1 \right) =
\frac{1}{2\sqrt{3\pi}} k_\pi\sqrt{\frac{k_\pi}{M_R}} 
\sqrt{E_N + M_N} \frac{g_{RN\pi}}{M_\pi},
\end{align}
%EQUATION<<<
for $j^P=3/2^+$ resonance, and
%EQUATION>>>
\begin{align}
G\left( 2 \right) = -
\frac{1}{2\sqrt{3\pi}} k_\pi \sqrt{\frac{k_\pi}{M_R}} 
\sqrt{E_N - M_N} \frac{g_{RN\pi}}{M_\pi},
\end{align}
%EQUATION<<<
for $j^P=3/2^-$ resonance.
%EQUATION<<<

\subsubsection{Spin-$5/2$ resonance}

The $RN\pi$ interaction Lagrangian for resonances of $j \geq 3/2$ can be obtained by
taking more derivatives on the pion field.
In the case of the resonance with $j^P = 5/2^\pm$, it can be written as
%EQUATION>>>
\begin{align}
\mathcal{L}_{RN\pi}^{(5/2\pm)} =&
i\, \frac{g_{RN\pi}^{}}{M_\pi^2}\, \bar{N}\, \Gamma^{(\pm)}
\partial_\mu \partial_\nu \pi\, R^{\mu\nu}
\cr   &
\pm i\, \frac{g_{RN\pi}^{}}{M_\pi^2}\, \bar{R}^{\mu\nu}\, \Gamma^{(\pm)}
\partial_\mu \partial_\nu \pi\, N,
\label{lag:pi5}
\end{align}
%EQUATION<<
which leads to
%EQUATION>>>
\begin{align}
\Gamma({\textstyle\frac52}^\pm \to N\pi) = \frac{2}{15} \frac{g_{RN\pi}^2}{4\pi}
\frac{k_\pi^5}{M_R M_\pi^4} (E_N \mp M_N).
\end{align}
%EQUATION<<<
This is consistent with the expression given in Ref.~\cite{Riska:2000gd}.

The obtained partial-wave decay amplitude is given by
%EQUATION>>>
\begin{align}
G \left( 3 \right) = -
\frac{1}{\sqrt{30\pi}} k_\pi^2 \sqrt{\frac{k_\pi}{M_R}} 
\sqrt{E_N - M_N} \frac{g_{RN\pi}}{M_\pi^2},
\end{align}
%EQUATION<<<
for $j^P=5/2^+$ resonance, and
%EQUATION>>>
\begin{align}
G \left( 2 \right) =
\frac{1}{\sqrt{30\pi}} k_\pi^2 \sqrt{\frac{k_\pi}{M_R}} 
\sqrt{E_N + M_N} \frac{g_{RN\pi}}{M_\pi^2},
\end{align}
%EQUATION<<<
for $j^P=5/2^-$ resonance.
%EQUATION<<<

\subsubsection{Spin-$7/2$ resonance}

In this case, we have
%EQUATION>>>
\begin{align}
\mathcal{L}_{RN\pi}^{(7/2\pm)} =&
- \frac{g_{RN\pi}^{}}{M_\pi^3}\, \bar{N} \, \Gamma^{(\mp)}
\partial_\mu \partial_\nu \partial_\alpha \pi\, R^{\mu\nu\alpha}
\cr   &
\mp \frac{g_{RN\pi}^{}}{M_\pi^3}\, \bar{R}^{\mu\nu\alpha}\, \Gamma^{(\mp)}
\partial_\mu \partial_\nu \partial_\alpha \pi\, N,
\label{lag:pi7}
\end{align}
%EQUATION>>>
which leads to
%EQUATION>>>
\begin{align}
\Gamma({\textstyle\frac72}^\pm \to N\pi) = \frac{2}{35} \frac{g_{RN\pi}^2}{4\pi} 
\frac{k_\pi^{7}}{M_R M_\pi^6} (E_N \pm M_N),
\end{align}
%EQUATION<<<
and
%EQUATION>>>
\begin{align}
G\left( 3 \right) =&
\sqrt{\frac{1}{70\pi}} k_\pi^3 \sqrt{\frac{k_\pi}{M_R}} 
\sqrt{E_N + M_N} \frac{g_{RN\pi}}{M_\pi^3},
\end{align}
%EQUATION<<<
for $j^P=7/2^+$ resonance, and
%EQUATION>>>
\begin{align}
G\left( 4 \right) =& -
\sqrt{\frac{1}{70\pi}} k_\pi^3 \sqrt{\frac{k_\pi}{M_R}} 
\sqrt{E_N - M_N} \frac{g_{RN\pi}}{M_\pi^3},
\end{align}
%EQUATION<<<
for $j^P=7/2^-$ resonance.
%EQUATION<<

\subsubsection{Spin-$9/2$ resonance}

In this case, we have
%EQUATION>>>
\begin{align}
\mathcal{L}_{RN\pi}^{(9/2\pm)} =&
-i\, \frac{g_{RN\pi}^{}}{M_\pi^4}\, \bar{N} \, \Gamma^{(\pm)}
\partial_\mu \partial_\nu \partial_\alpha \partial_\beta \pi\, R^{\mu\nu\alpha\beta}
\cr   &
\mp i \frac{g_{RN\pi}^{}}{M_\pi^4}\, \bar{R}^{\mu\nu\alpha\beta}\, \Gamma^{(\pm)}
\partial_\mu \partial_\nu \partial_\alpha \partial_\beta \pi\, N,
\label{lag:pi9}
\end{align}
%EQUATION<<<
which leads to
%EQUATION>>>
\begin{align}
\Gamma({\textstyle\frac92}^\pm \to N\pi) = \frac{8}{315} \frac{g_{RN\pi}^2}{4\pi} 
\frac{k_\pi^{9}}{M_R M_\pi^8} (E_N \mp M_N),
\end{align}
%EQUATION<<<
and
%EQUATION>>>
\begin{align}
G\left( 5 \right) = -
\sqrt{\frac{2}{315\pi}} k_\pi^4 \sqrt{\frac{k_\pi}{M_R}} 
\sqrt{E_N - M_N} \frac{g_{RN\pi}}{M_\pi^4},
\end{align}
%EQUATION<<<
for $j^P=9/2^+$ resonance, and
%EQUATION>>>
\begin{align}
G\left( 4 \right) =
\sqrt{\frac{2}{315\pi}} k_\pi^4 \sqrt{\frac{k_\pi}{M_R}} 
\sqrt{E_N + M_N} \frac{g_{RN\pi}}{M_\pi^4},
\end{align}
%EQUATION<<<
for $j^P=9/2^-$ resonance.

\subsubsection{Summary}

Thus, the general form of the effective Lagrangian for the $RN\pi$ interaction can
be written as
%EQUATION>>>
\begin{align}
\mathcal{L}_{RN\pi} = \frac{g_{RN\pi}^{}}{M_\pi^{n-1}}
\bar{N} \partial_{\mu_1^{}} \cdots \partial_{\mu_{n-1}^{}} \pi \left[ i \Gamma^{(\pm)}
\right] R^{\mu_1^{} \cdots \mu_{n-1}^{}} + \mbox{H.c.},
\end{align}
%EQUATION<<<
where $R$ stands for a spin-$j$ resonance with $j = n - 1/2$ and $N$ is the
spin-$1/2$ field.
The insertion of the factor $[ i \Gamma^{(\pm)} ]$ should be made depending on the
spin-parity of the resonance and the pion mass $M_\pi$ is introduced to make the
coupling constant $g_{RN\pi}^{}$ dimensionless.

The decay widths for $j^P \to 0^- + 1/2^+$ obtained above are summarized below.
%EQUATION>>>
\begin{align}
\Gamma(\textstyle\frac12^\pm \to N\pi) =& \frac{g^2}{4\pi}
\frac{k_\pi^{}}{M_R} (E_N \mp M_N),
\cr
\Gamma(\textstyle\frac32^\pm \to N\pi) =& \frac{1}{3} \frac{g^2}{4\pi}
\frac{k_\pi^{3}}{M_R M_\pi^2} (E_N \pm M_N),
\cr
\Gamma(\textstyle\frac52^\pm \to N\pi) =& \frac{2}{15} \frac{g^2}{4\pi} 
\frac{k_\pi^{5}}{M_R M_\pi^4} (E_N \mp M_N),
\cr
\Gamma(\textstyle\frac72^\pm \to N\pi) =& \frac{2}{35} \frac{g^2}{4\pi} 
\frac{k_\pi^{7}}{M_R M_\pi^6} (E_N \pm M_N),
\cr
\Gamma(\textstyle\frac92^\pm \to N\pi) =& \frac{8}{315} \frac{g^2}{4\pi} 
\frac{k_\pi^{9}}{M_R M_\pi^8} (E_N \mp M_N).
\end{align}
%EQUATION<<<

Our results are consistent with the formula of Ref.~\cite{Rushbrooke:1966zz} for
the width of the $R \to N\pi$ decay, which reads
%EQUATION>>>
\begin{align}
\Gamma(R \to N\pi) = \frac{g^2}{4\pi} \frac{2^n (n!)^2}{n(2n)!}
\frac{k_\pi^{2n-1}}{M_R M_\pi^{2(n-1)}} (E_N \pm M_N),
\end{align}
%EQUATION<<<
for $(-1)^nP_j = \pm 1$, where $P_j$ is the parity of the spin-$j$
resonance $R$ and $n=j+1/2$.

%TABLE>>>
\begin{table*}[ht]
\centering
\begin{tabular}{c|cc|cc} \hline\hline
& \multicolumn{2}{c|}{$(-1)^\gamma = +1$} & \multicolumn{2}{c}{$(-1)^\gamma = -1$} \\ 
& $J_0$ & $J_{+1}$ & $J_0$ & $J_{+1}$ \\ \hline
Nonidentical fermions ($s' = s$) & $j_{\rm min}^{} + \frac12$ & $2 j_{\rm min}$ &
$j_{\rm min}^{} + \frac12$ & $2 j_{\rm min}$ \\
\phantom{Nonidentical fermions} ($s' \neq s$) & $j_{\rm min}^{} + \frac12$ &
$2 j_{\rm min} + 1$ &
 $j_{\rm min}^{} + \frac12$ & $2 j_{\rm min} + 1$ \\
Nonidentical bosons ($s' = s$) & $j_{\rm min}^{} + 1$ & $2 j_{\rm min}^{}$ &
$j_{\rm min}^{}$ & $2 j_{\rm min}$ \\
\phantom{Nonidentical bosons} ($s' \neq s$) & $j_{\rm min}^{} + 1$ & $2 j_{\rm min}^{}
+ 1$ & $j_{\rm min}^{}$ & $2 j_{\rm min} + 1$ \\
Identical fermions($s' = s$)& --- & --- & $j_{\rm min}^{} + \frac12$ & $j_{\rm min}^{}
+ \frac12$  \\
Identical bosons ($s' = s$)& $j_{\rm min}^{} + 1$ & $j_{\rm min}^{}$ & --- & --- \\
\hline\hline
\end{tabular}
\caption{Number of independent couplings for the vector current of a spin-$s'$
particle into a spin-$s$ particle transition~\cite{Durand:1962zza}.
Here $\gamma = s+s'+P$, where $(-1)^P$ is the relative parity of
the initial and final states whose spins are $s'$ and $s$, respectively, and
$j_{\rm min} = \min(s,s')$. See Ref.~\cite{Durand:1962zza} for the details.}
\label{TAB02}
\end{table*}
%TABLE<<<

\subsection{Vector meson couplings}

\begin{figure}[h]
\centering
%\vskip -0.4cm
\includegraphics[width=0.25\textwidth,angle=0,clip]{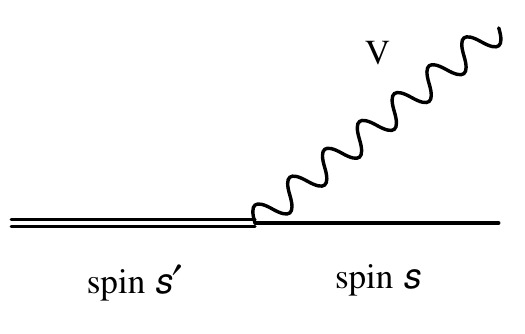}
\caption{Vertex of the vector-meson decay of a spin-$s'$ baryon into a spin-$s$ baryon.}
\label{FIG02}
\end{figure}

We now consider the decays of a spin-$s'$ baryon into the vector meson and
spin-$s$ baryon channel, as shown in Fig.~\ref{FIG02}.
The number of independent couplings is estimated in a more sophisticated way and
we quote the results of Ref.~\cite{Durand:1962zza} in Table~\ref{TAB02}.
Here $J_0$ is the time component of the vector current and $J_\pm$ is its space
component, $J_{\pm 1} = \mp (J_x \pm i J_y)/\sqrt2$.
The total number of independent couplings is the sum of the numbers for $J_0$ and
$J_{+1}$.

%FIGURE>>>
%FIGURE<<<
Table~\ref{TAB02} indicates that, in our case of $R \to NV$ decay, there are 3
independent couplings for $j^P \to 1^- +  1/2^+$ with $j \ge 3/2$, while the
$1/2^P \to 1^- + 1/2^+$ decay has 2 independent terms.

The helicity amplitudes for the decay of $R \to NV$ are related to the decay width as
%EQUATION>>>
\begin{align}
\Gamma(R \to NV) =& \frac{k_V^2}{\pi} \frac{2M_N}{(2j+1) M_R}
\cr
& \hskip -0.3cm \times
\left\{ \left| A_{1/2} \right|^2 + \left| S_{1/2} \right|^2 + \left| A_{3/2} \right|^2
\right\},
\end{align}
%EQUATION<<<
where $k_V =|{\bf k}_V|$ is the magnitude of the three-momentum of the vector meson
in the rest frame of the resonance
%EQUATION>>>
\begin{align}
k_V = \frac{1}{2M_R} 
\sqrt{[M_R^2 - (M_N+M_V)^2][M_R^2 - (M_N-M_V)^2]}.
\end{align}
%EQUATION<<<

The helicity amplitudes are then obtained as
%EQUATION>>>
\begin{align}
A_\lambda^{(j\pm)} =& \frac{1}{\sqrt{8M_N M_R k_V}} \frac{2j+1}{4\pi}
\cr   & \times
\int d \cos\theta\, d\phi \, e^{-i(m-\lambda)\phi} d^j_{\lambda m}(\theta)
\cr   & \times
\langle \textbf{k}_V, \lambda_V,\lambda_N \mid -i \mathcal{M}_{R \to NV} \mid j m
\rangle,
\label{eq:HA}
\end{align}
%EQUATION<<<
where $\mathcal{M}_{R \to NV}$ is the transition amplitude.
The lower index denotes the helicity of the final $N V$ state, i.e., $\lambda =
\lambda_V - \lambda_N$, where we use an abbraviation as follows:
$A_{1/2}:1 - \textstyle\frac12$,
$S_{1/2}:0 +  \textstyle\frac12$,
$A_{3/2}:1 + \textstyle\frac12$.

The details of the interaction Lagrangians and the relevant decay widths and helicity 
amplitudes are given below.

%------------------------------>>> widetext
\begin{widetext}

\subsubsection{Spin-$1/2$ resonance}

In this case, we have 2 independent terms in the interaction Lagrangian, which
reads~\cite{Tsushima:2003fs}
%EQUATION>>>
\begin{align}
\mathcal{L}^{(1/2\pm)}_{RNV} =&
- \frac{1}{2M_N} \bar{N}
\left[ g_2^{} \left( \pm \frac{\Gamma_\mu^{(\mp)}
\partial^2}{M_{R} \mp M_N} -i \Gamma^{(\mp)} \partial_\mu \right)
- g_{1}^{} \Gamma^{(\mp)} \sigma_{\mu\nu}^{}
\partial^\nu \right] V^\mu R
\cr   &
- \frac{1}{2M_N} \bar{R}
\left[ g_2^{} \left( \pm \frac{\Gamma_\mu^{(\mp)}
\partial^2}{M_{R} \mp M_N} \pm i \Gamma^{(\mp)} \partial_\mu \right)
\mp g_{1}^{}W_ \Gamma^{(\mp)} \sigma_{\mu\nu}^{}
\partial^\nu \right] V^\mu N,
\label{lag:vm1}
\end{align}
%EQUATION>>>
where $V^\mu$ is the vector-meson field.
Note that the $g_2^{}$ term contains $\partial^2 V^\mu$, which is required to
satisfy the gauge invariance condition.

This leads to the decay width as
%EQUATION>>>
\begin{align}
\Gamma({\textstyle\frac12}^\pm \to NV) =&
\frac{1}{16\pi} \frac{k_V(E_N\mp M_N)}{M_N^2 M_R}
\Biggl\{ g_1^2 \left[ 2 (M_R \pm M_N)^2 + M_V^2 \right]
- 6 g_1^{} g_2^{} \frac{M_V^2}{(M_R \mp M_N)^2} (M_R^2 - M_N^2)
\cr   &
\hspace{9em}
+ g_2^2 \frac{M_V^2}{(M_R \mp M_N)^2} 
\left[ (M_R \pm M_N)^2 + 2 M_V^2 \right] \Biggr\}.
\end{align}
%EQUATION<<<
Note that all $g_2^{}$ terms contain the vector meson mass $M_V$.

\subsubsection{Spin $\geq 3/2$ resonances}

The effective Lagrangian for the $RNV$ interaction for the resonance of spin $j \ge
3/2$ can be constructed by using the field strength tensor of the vector meson
$V_{\mu\nu} = \partial_\mu V_\nu - \partial_\nu V_\mu$.
As discussed above, there are 3 independent terms in this case.
We have
%EQUATION>>>
\begin{align}
\mathcal{L}_{RNV}^{(3/2\pm)} =&
\left[
- \frac{ig_1^{}}{2M_N} \bar{N} \Gamma_\nu^{(\pm)}
- \frac{g_2}{(2M_N)^2} \partial_\nu \bar{N} \Gamma^{(\pm)}
+ \frac{g_3}{(2M_N)^2} \bar{N} \Gamma^{(\pm)} \partial_\nu
\right]
V^{\mu\nu} R_\mu + \mbox{H.c.},
\cr
\mathcal{L}_{RNV}^{(5/2\pm)} =&
\left[
\frac{g_1^{}}{(2M_N)^2} \bar{N} \Gamma_\nu^{(\mp)}
- \frac{ig_2^{}}{(2M_N)^3} \partial_\nu \bar{N} \Gamma^{(\mp)}
+ \frac{ig_3^{}}{(2M_N)^3} \bar{N} \Gamma^{(\mp)} \partial_\nu
\right]
\partial^\alpha V^{\mu\nu} R_{\mu\alpha} + \mbox{H.c.},
\cr
\mathcal{L}_{RNV}^{(7/2\pm)} =&
\left[
\frac{ig_1^{}}{(2M_N)^3} \bar{N} \Gamma_\nu^{(\pm)} 
+ \frac{g_2^{}}{(2M_N)^4} \partial_\nu \bar{N} \Gamma^{(\pm)}
- \frac{g_3^{}}{(2M_N)^4} \bar{N} \Gamma^{(\pm)} \partial_\nu
\right]
\partial^{\alpha} \partial^{\beta} V^{\mu\nu} R_{\mu\alpha\beta} + \mbox{H.c.},
\cr
\mathcal{L}_{RNV}^{(9/2\pm)} =&
\left[
- \frac{g_1^{}}{(2M_N)^4} \bar{N} \Gamma_\nu^{(\mp)}
+ \frac{i g_2^{}}{(2M_N)^5} \partial_\nu \bar{N} \Gamma^{(\mp)}
- \frac{i g_3^{}}{(2M_N)^5} \bar{N} \Gamma^{(\mp)} \partial_\nu
\right]
\partial^{\alpha} \partial^{\beta} \partial^{\rho} V^{\mu\nu} R_{\mu\alpha\beta\rho}
+ \mbox{H.c.}.
\end{align}
%EQUATION<<<

Then the decay widths are obtained as follows

\textbullet\ spin-$3/2$
\begin{subequations}
%EQUATION>>>
\begin{align}
\Gamma({\textstyle\frac32}^\pm \to NV) =&
\frac{1}{12\pi} \frac{k_V}{M_R} (E_N\mp M_N)
\cr   &   \times
\biggl\{
\bar{g}_1^2 \left[ 2 E_N (E_N \pm M_N) + (M_R \pm M_N)^2 + 2M_V^2 \right]
+ \bar{g}_2^2 \left[ E_N^2 (2M_R^2 + M_V^2) - 2 M_N^2 (M_R^2 - M_V^2) \right]
\cr   &
+ \bar{g}_3^2 M_V^2 ( E_N^2 - M_N^2 + 3 M_V^2) \mp 2 \bar{g}_1^{} \bar{g}_2^{}
  \left[ \frac{E_N}{2} ( 3 M_R^2 + M_N^2 \pm 2 M_N M_R + 3 M_V^2) - M_N^2 (2M_R \pm M_N)
\right]
\cr   &
+ 2 \bar{g}_2^{} \bar{g}_3^{} M_V^2 (E_N^2 + 2M_N^2 - 3E_N M_R)
\pm 2 \bar{g}_3^{} \bar{g}_1^{} M_V^2 ( 3M_R -2 E_N \pm M_N)
\biggr\}.
\end{align}
%EQUATION<<<
\end{subequations}

\textbullet\ spin-$5/2$
\begin{subequations}
%EQUATION>>>
\begin{align}
\Gamma({\textstyle\frac52}^\pm \to NV) =&
\frac{1}{60\pi} \frac{k_V^3}{M_R} (E_N\pm M_N)
\cr   &   \times
\biggl\{
\bar{g}_1^2 \left[ 4 E_N (E_N \mp M_N) + (M_R \mp M_N)^2
+ 4M_V^2 \right]
+ \bar{g}_2^2 \left[ E_N^2 (3M_R^2 + 2M_V^2) - 3 M_N^2 (M_R^2 - M_V^2)
\right]
\cr   &
+ \bar{g}_3^2 M_V^2 ( 2E_N^2 - 2M_N^2 + 5 M_V^2)
\pm 2 \bar{g}_1 \bar{g}_2 \left[ E_N ( 2M_R^2 + M_N^2 \mp M_N M_R + 3
M_V^2)
- M_N^2 (3M_R \mp M_N) \right]
\cr   &
+ 2 \bar{g}_2 \bar{g}_3 M_V^2 (2E_N^2 + 3M_N^2 - 5E_N M_R)
\mp 2 \bar{g}_3 \bar{g}_1 M_V^2 ( 5M_R -4 E_N \mp M_N)
\biggr\}.
\end{align}
%EQUATION<<<
\end{subequations}

\textbullet\ spin-$7/2$
\begin{subequations}
%EQUATION>>>
\begin{align}
\Gamma({\textstyle\frac72}^\pm \to NV) =&
\frac{1}{210\pi} \frac{k_V^5}{M_R} (E_N\mp M_N)
\cr   & \times
\biggl\{
\bar{g}_1^2 \left[ 6 E_N (E_N \pm M_N) + (M_R \pm M_N)^2 + 6M_V^2 \right]
+ \bar{g}_2^2 \left[ E_N^2 (4M_R^2 + 3M_V^2) - 4 M_N^2 (M_R^2 - M_V^2) \right]
\cr   &
+ \bar{g}_3^2 M_V^2 ( 3E_N^2 - 3M_N^2 + 7 M_V^2)
\mp 2 \bar{g}_1^{} \bar{g}_2^{} \left[ \frac{E_N}{2} ( 5M_R^2 + 3 M_N^2 \pm 2 M_N M_R
+ 9 M_V^2) - M_N^2 (4M_R \pm M_N) \right]
\cr   &
+ 2 \bar{g}_2^{} \bar{g}_3^{} M_V^2 (3E_N^2 + 4M_N^2 - 7E_N M_R)
\pm 2 \bar{g}_3^{} \bar{g}_1^{} M_V^2 ( 7M_R -6 E_N \pm M_N)
\biggr\}.
\end{align}
%EQUATION<<<
\end{subequations}

\textbullet\ spin-$9/2$
\begin{subequations}
%EQUATION>>>
\begin{align}
\Gamma({\textstyle\frac92}^\pm \to NV) =&
\frac{1}{630\pi} \frac{k_V^7}{M_R} (E_N \pm M_N)
\cr   & \times
\biggl\{
\bar{g}_1^2 \left[ 8 E_N (E_N \mp M_N) + (M_R \mp M_N)^2 + 8M_V^2 \right]
+ \bar{g}_2^2 \left[ E_N^2 (5M_R^2 + 4M_V^2) - 5 M_N^2 (M_R^2 - M_V^2) \right]
\cr   &
+ \bar{g}_3^2 M_V^2 ( 4E_N^2 - 4M_N^2 + 9 M_V^2)
\pm 2 \bar{g}_1^{} \bar{g}_2^{} \left[ E_N ( 3M_R^2 + 2 M_N^2 
\mp M_N M_R + 6 M_V^2) - M_N^2 (5M_R \mp M_N) \right]
\cr   &
+ 2 \bar{g}_2^{} \bar{g}_3^{} M_V^2 (4E_N^2 + 5M_N^2 - 9E_N M_R)
\mp 2 \bar{g}_3^{} \bar{g}_1^{} M_V^2 ( 9M_R - 8E_N \mp M_N)
\biggr\}.
\end{align}
%EQUATION<<<
\end{subequations}
For $n = j +1/2$, we have
%EQUATION>>>
\begin{align}
\bar{g}_1 = \frac{g_1^{}}{(2M_N)^{n-1}}, \,
\bar{g}_2 = \frac{g_2^{}}{(2M_N)^n}, \,
\bar{g}_3 = \frac{g_3^{}}{(2M_N)^n}.
\end{align}
%EQUATION<<<

\subsubsection{Helicity amplitudes}

The helicity amplitudes derived from Eq.~(\ref{eq:HA}) read

\textbullet\ spin-$1/2$
\begin{subequations}
%EQUATION>>>
\begin{align}
A_{1/2}^{(1/2\pm)} =&
\mp \frac{1}{2\sqrt2} \frac{\sqrt{E_N \mp M_N}}{M_N\sqrt{k_V M_N}}
\left[ (M_R \pm M_N) g_1^{} - \frac{M_V^2}{M_R \mp M_N} g_2^{} \right],
\cr
S_{1/2}^{(1/2\pm)} =&
- \frac{M_V}{4M_N} \sqrt{\frac{E_N \mp M_N}{k_V M_N}}
\left[ g_1^{} - \frac{M_R \pm M_N}{M_R \mp M_N} g_2^{} \right].
\end{align}
%EQUATION>>>
\end{subequations}

\textbullet\ spin-$3/2$
\begin{subequations}
%EQUATION>>>
\begin{align}
A_{1/2}^{(3/2\pm)} =&
\mp \frac{1}{4\sqrt3} \frac{\sqrt{E_N \mp M_N}}{M_N\sqrt{k_V M_N}} 
\left\{ \frac{g_1^{}}{M_R} \left[ M_N (M_R \pm M_N) \mp M_V^2 \right] + 
\frac{g_2^{}}{4M_N} (M_R^2 - M_N^2 - M_V^2) - 
\frac{g_3^{}}{2M_N} M_V^2 \right\},
\cr
S_{1/2}^{(3/2\pm)} =&
\mp \frac{1}{2\sqrt6} \frac{M_V \sqrt{E_N \mp M_N}}{M_N \sqrt{k_V M_N}} 
\left\{ g_1^{} \mp \frac{g_2^{}}{4M_N M_R} (M_R^2 + M_N^2 - M_V^2) \pm 
\frac{g_3^{}}{4 M_N M_R} (M_R^2 - M_N^2 + M_V^2) \right\},
\cr
A_{3/2}^{(3/2\pm)} =&
\mp \frac{1}{4} \frac{\sqrt{E_N \mp M_N}}{M_N \sqrt{k_V M_N}}
\left\{ g_1^{} (M_R \pm M_N) \mp 
\frac{g_2^{}}{4M_N} (M_R^2 - M_N^2 - M_V^2) \pm 
\frac{g_3^{}}{2M_N} M_V^2 \right\}.
\end{align}
%EQUATION>>>
\end{subequations}

\textbullet\ spin-$5/2$
\begin{subequations}
%EQUATION>>>
\begin{align}
A_{1/2}^{(5/2\pm)} = \frac{\sqrt{3}}{\sqrt{20}} \frac{k_\gamma}{M_N} A_{1/2}^{(3/2\mp)},\,\,\,
S_{1/2}^{(5/2\pm)} = \frac{\sqrt{3}}{\sqrt{20}} \frac{k_\gamma}{M_N} S_{1/2}^{(3/2\mp)},\,\,\,
A_{3/2}^{(5/2\pm)} = \frac{1}{\sqrt{10}} \frac{k_\gamma}{M_N} A_{3/2}^{(3/2\mp)}.
\end{align}
%EQUATION<<<
\end{subequations}

\textbullet\ spin-$7/2$
\begin{subequations}
%EQUATION>>>
\begin{align}
A_{1/2}^{(7/2\pm)} = \frac{1}{\sqrt7} \frac{k_\gamma}{M_N} A_{1/2}^{(5/2\mp)},\,\,\,
S_{1/2}^{(7/2\pm)} = \frac{1}{\sqrt7} \frac{k_\gamma}{M_N} S_{1/2}^{(5/2\mp)},\,\,\,
A_{3/2}^{(7/2\pm)} = \frac{\sqrt{5}}{\sqrt{42}} \frac{k_\gamma}{M_N} A_{3/2}^{(5/2\mp)}.
\end{align}
%EQUATION<<<
\end{subequations}

\textbullet\ spin-$9/2$
\begin{subequations}
%EQUATION>>>
\begin{align}
A_{1/2}^{(9/2\pm)} = \frac{\sqrt{5}}{6} \frac{k_\gamma}{M_N} A_{1/2}^{(7/2\mp)},\,\,\,
S_{1/2}^{(9/2\pm)} = \frac{\sqrt{5}}{6} \frac{k_\gamma}{M_N} S_{1/2}^{(7/2\mp)},\,\,\,
A_{3/2}^{(9/2\pm)} = \frac{1}{\sqrt{8}} \frac{k_\gamma}{M_N} A_{3/2}^{(7/2\mp)}.
\end{align}
%EQUATION<<<
\end{subequations}

% ------------------------------>>> Widetext
\end{widetext}

%-------------------------------------------------------------------------------------
\section{Couplings of baryon resonances into a photon and a
spin-$1/2$ baryon} \label{Sec:IV}
%-------------------------------------------------------------------------------------

The analysis obtained in Sec.~\ref{Sec:III} can be directly applied to the
photo-decay of resonances into the nucleon.
In this case, since the vector current is the real photon, we have an additional
condition, i.e., the absence of the helicity-0 state, which reduces the number of
independent couplings by one.
In fact, the $g_3^{}$ terms in Sec.~\ref{Sec:III} are not present in the case of real
photon, and the results can be readily obtained from those for the vector meson case.

The interaction Lagrangians read
%EQUATION>>>
\begin{align}
\mathcal{L}_{RN\gamma}^{(1/2\pm)} =&
\frac{ef_1}{2M_N} \bar{N} \Gamma^{(\mp)} \sigma_{\mu\nu} \partial^\nu A^\mu R
\cr   &
\pm \frac{ef_1}{2M_N} \bar{R} \Gamma^{(\mp)} \sigma_{\mu\nu} \partial^\nu A^\mu N,
\cr
\mathcal{L}_{RN\gamma}^{(3/2\pm)} =&
\left[
- \frac{ief_1}{2M_N} \bar{N} \Gamma_\nu^{(\pm)}
- \frac{ef_2}{(2M_N)^2} \partial_\nu \bar{N} \Gamma^{(\pm)}
\right]
\cr   & \times F^{\mu\nu} R_\mu + \mbox{H.c.},
\cr
\mathcal{L}_{RN\gamma}^{(5/2\pm)} =&
\left[
\frac{ef_1}{(2M_N)^2} \bar{N} \Gamma_\nu^{(\mp)}
- \frac{ief_2}{(2M_N)^3} \partial_\nu \bar{N} \Gamma^{(\mp)}
\right]
\cr   & \times \partial^\alpha F^{\mu\nu} R_{\mu\alpha}
+ \mbox{H.c.},
\cr
\mathcal{L}_{RN\gamma}^{(7/2\pm)} =&
\left[
\frac{ief_1}{(2M_N)^3} \bar{N} \Gamma_\nu^{(\pm)}
+ \frac{ef_2}{(2M_N)^4} \partial_\nu \bar{N} \Gamma^{(\pm)}
\right]
\cr   & \times \partial^{\alpha} \partial^{\beta} F^{\mu\nu} R_{\mu\alpha\beta}
+ \mbox{H.c.},
\cr
\mathcal{L}_{RN\gamma}^{(9/2\pm)} =&
\left[
- \frac{ef_1}{(2M_N)^4} \bar{N} \Gamma_\nu^{(\mp)}
+ \frac{ief_2}{(2M_N)^5} \partial_\nu \bar{N} \Gamma^{(\mp)}
\right]
\cr & \times \partial^{\alpha} \partial^{\beta} \partial^\rho F^{\mu\nu} R_{\mu\alpha\beta\rho}
+ \mbox{H.c.},
\end{align}
%EQUATION<<<
where $F_{\mu\nu} = \partial_\mu A_\nu - \partial_\nu A_\mu$ and $A^\mu$ the photon field.

The obtained decay widths are given as follows~\cite{Oh:2011}

\textbullet\ spin-$1/2$
\begin{subequations}
%EQUATION>>>
\begin{align}
\Gamma({\textstyle\frac12}^\pm \to N\gamma) = \alpha_{\rm em} f_1^2
\frac{k_\gamma^3}{M_N^2}.
\end{align}
%EQUATION<<<
\end{subequations}

\textbullet\ spin-$3/2$
\begin{subequations}
%EQUATION>>>
\begin{align}
\Gamma({\textstyle\frac32}^\pm \to N\gamma) &= \frac{\alpha_{\rm em}}{12}
\frac{k_\gamma^3}{M_R^2}
\cr   & \hspace{-3em} \times
\biggl\{ \left[ \bar{f}_1 (3 M_R \pm M_N) \mp \bar{f}_2 M_R (M_R \mp M_N) \right]^2
\cr   & \hspace{-3em}
+ 3 \left[ \bar{f}_1 \mp \bar{f}_2 M_R \right]^2 (M_R \mp M_N)^2 \biggr\}.
\end{align}
%EQUATION<<<
\end{subequations}

\textbullet\ spin-$5/2$

\begin{subequations}
%EQUATION>>>
\begin{align}
\Gamma({\textstyle\frac52}^\pm \to N\gamma) &= \frac{\alpha_{\rm em}}{90}
\frac{k_\gamma^5}{M_R^2}
\cr   & \hspace{-3em} \times
\biggl\{ \left[ \bar{f}_1 (4M_R \mp 2M_N) \pm \bar{f}_2 M_R (M_R \pm M_N) \right]^2
\cr   & \hspace{-3em}
+ 8 \left[ \bar{f}_1 \pm \bar{f}_2 M_R \right]^2 (M_R \pm M_N)^2 \biggr\}.
\end{align}
%EQUATION<<<
\end{subequations}

\textbullet\ spin-$7/2$
\begin{subequations}
%EQUATION>>>
\begin{align}
\Gamma({\textstyle\frac72}^\pm \to N\gamma) &= \frac{\alpha_{\rm em}}{420}
\frac{k_\gamma^7}{M_R^2}
\cr   & \hspace{-3em} \times
\biggl\{ \left[ \bar{f}_1 (5M_R \pm 3M_N) \mp \bar{f}_2 M_R (M_R \mp M_N) \right]^2
\cr   & \hspace{-3em}
+ 15 \left[ \bar{f}_1 \mp \bar{f}_2 M_R \right]^2 (M_R \mp M_N)^2 \biggr\}.
\end{align}
%EQUATION<<<
\end{subequations}

\textbullet\ spin-$9/2$
\begin{subequations}
%EQUATION>>>
\begin{align}
\Gamma({\textstyle\frac92}^\pm \to N\gamma) &= \frac{\alpha_{\rm em}}{1575}
\frac{k_\gamma^9}{M_R^2}
\cr   & \hspace{-3em} \times
\biggl\{ \left[ \bar{f}_1 (6M_R \mp 4M_N) \pm \bar{f}_2 M_R (M_R \pm M_N) \right]^2
\cr   & \hspace{-3em}
+ 24 \left[ \bar{f}_1 \pm \bar{f}_2 M_R \right]^2 (M_R \pm M_N)^2 \biggr\},
\end{align}
%EQUATION<<<
\end{subequations}
where
%EQUATION>>>
\begin{align}
& \alpha_{\rm em}^{} = \frac{e^2}{4\pi},\,\,\,
  k_\gamma = \frac{1}{2M_R} \left( M_R^2 - M_N^2 \right), \,\,\,
\cr
& \bar{f}_1 = \frac{f_1}{(2M_N)^{n-1}}, \,\,\,
  \bar{f}_2 = \frac{f_2}{(2M_N)^n},
\end{align}
%EQUATION<<<
for $n = j + 1/2$.

The decay width for $R \to N\gamma$ is related to the helicity amplitudes of the
photo-transition as
%EQUATION>>>
\begin{align}
\Gamma (R \to N\gamma) = \frac{k_\gamma^2}{\pi} \frac{2M_N}{(2j+1) M_R}
\left\{ |A_{1/2}|^2 + |A_{3/2}|^2 \right\}.
\end{align}
%EQUATION<<<
Therefore, we have~\cite{Shklyar:2004ba,Oh:2007jd,Oh:2011}

\textbullet\ spin-$1/2$
\begin{subequations}
%EQUATION>>>
\begin{align}
A_{1/2}^{(1/2\pm)} =
\mp \frac{e f_1}{2M_N} \sqrt{\frac{k_\gamma M_R}{M_N}}.
\end{align}
%EQUATION<<<
\end{subequations}

\textbullet\ spin-$3/2$
\begin{subequations}
%EQUATION>>>
\begin{align}
A_{1/2}^{(3/2\pm)}
=& \mp \frac{e\sqrt6}{12} \sqrt{\frac{k_\gamma}{M_N M_R}}
\cr   &   \times
\left[ f_1 + \frac{f_2}{4M_N^2} M_R (M_R \mp M_N) \right],
\cr
A_{3/2}^{(3/2\pm)}
=& \mp \frac{e\sqrt2}{4M_N} \sqrt{\frac{k_\gamma M_R}{M_N}}
\cr   &    \times
\left[ f_1 \mp \frac{f_2}{4M_N} (M_R \mp M_N) \right].
\end{align}
%EQUATION<<<
\end{subequations}

\textbullet\ spin-$5/2$
\begin{subequations}
%EQUATION>>>
\begin{align}
A_{1/2}^{(5/2\pm)}
=& \pm \frac{e}{4\sqrt{10}} \frac{k_\gamma}{M_N} \sqrt{\frac{k_\gamma}{M_N M_R}}
\cr   &   \times
\left[ f_1 + \frac{f_2}{4M_N^2} M_R (M_R \pm M_N) \right],
\cr
A_{3/2}^{(5/2\pm)}
=& \pm \frac{e}{4\sqrt5} \frac{k_\gamma}{M_N^2} \sqrt{\frac{k_\gamma M_R}{M_N}}
\cr   &   \times
\left[ f_1 \pm \frac{f_2}{4M_N} (M_R \pm M_N) \right].
\end{align}
%EQUATION<<<
\end{subequations}

\textbullet\ spin-$7/2$
\begin{subequations}
%EQUATION>>>
\begin{align}
A_{1/2}^{(7/2\pm)}
=& \mp \frac{e\sqrt{70}}{280} \frac{k_\gamma^2}{M_N^2} \sqrt{\frac{k_\gamma}{M_N M_R}}
\cr   &   \times
\left[ f_1 + \frac{f_2}{4M_N^2} M_R (M_R \mp M_N) \right],
\cr
A_{3/2}^{(7/2\pm)}
=& \mp \frac{e\sqrt{42}}{168} \frac{k_\gamma^2}{M_N^3} \sqrt{\frac{k_\gamma M_R}{M_N}}
\cr   &   \times
\left[ f_1 \mp \frac{f_2}{4M_N} (M_R \mp M_N) \right].
\end{align}
%EQUATION<<<
\end{subequations}

\textbullet\ spin-$9/2$
\begin{subequations}
%EQUATION>>>
\begin{align}
A_{1/2}^{(9/2\pm)}
=& \pm \frac{e\sqrt{14}}{336} \frac{k_\gamma^3}{M_N^3} \sqrt{\frac{k_\gamma}{M_N M_R}}
\cr   &   \times
\left[ f_1 + \frac{f_2}{4M_N^2} M_R (M_R \pm M_N) \right],
\cr
A_{3/2}^{(9/2\pm)}
=& \pm \frac{e\sqrt{21}}{336} \frac{k_\gamma^3}{M_N^4} \sqrt{\frac{k_\gamma M_R}{M_N}}
\cr   &   \times
\left[ f_1 \pm \frac{f_2}{4M_N} (M_R \pm M_N) \right].
\end{align}
%EQUATION<<<
\end{subequations}

All of them reproduce the results for $\Gamma(R \to N\gamma)$ given before.
Note that the lower indices stand for the $N \gamma$ helicities
$\lambda = \lambda_\gamma - \lambda_N$, i.e.,
$A_{1/2}:1 - \textstyle\frac12$,
$A_{3/2}:1 + \textstyle\frac12$.

%-------------------------------------------------------------------------------------
\section{Couplings of baryon resonances into a meson and a
spin-$3/2$ baryon} \label{Sec:V}
%-------------------------------------------------------------------------------------

In this section, we construct the general form of the $R \Delta \pi$ and $R \Delta V$
interactions, where $\Delta$ denotes a spin-3/2 baryon.
The effective Lagrangians obtained for these interactions describe the decays of
$j^P \to 0^- + 3/2^+$ and $j^P \to 1^- + 3/2^+$.
As mentioned previously, we do \textit{not} consider the isospin factor.

\subsection{Pseudoscalar meson couplings}

For the decay of $j^P \to 0^- + 3/2^+$, the number of independent couplings is
indicated in Table~\ref{TAB01}.
There are only one coupling for $j = 1/2$ resonances and 2 independent terms
for resonances with $j \ge 3/2$.

The decay amplitude for $R \to \Delta \pi$ can be expressed as~\cite{Oh:2007jd}
%EQUATION>>>
\begin{align}
& \left < \pi({\bf q_\pi})\,\Delta(-{\bf q_\pi},m_f) | -i \mathcal{H}_\mathrm{int} |
R({\bf 0},m_j) \right >
\cr   \hspace{-1em} =&
4 \pi M_R \sqrt\frac{2}{q_\pi} \sum_{\ell,m_\ell}
\langle \ell\, m_\ell\, {\textstyle\frac{3}{2}}\, m_f | j \,m_j \rangle Y_{\ell,m_\ell} 
({\bf \hat {q_\pi}}) G(\ell),
\label{eq:DA2}
\end{align}
%EQUAITON<<<
where $\langle\ell\,m_\ell\,\frac{3}{2}\,m_f|j \,m_j \rangle$ and $Y_{\ell,m_\ell}
({\bf \hat {q_\pi}})$ are the Clebsch-Gordan coefficient and spherical harmonics,
respectively. 
$q_\pi=|{\bf q}_\pi|$ is the magnitude of the three-momentum of the pion in the rest
frame of the resonance,
%EQUATION>>>
\begin{align}
q_\pi = \frac{1}{2M_R}
\sqrt{[M_R^2 - (M_\Delta+M_\pi)^2][M_R^2 - (M_\Delta-M_\pi)^2]}.
\end{align}
%EQUATION<<<
The relative orbital angular momentum $\ell$ of the $\Delta \pi$ final state is
constrained by the spin-parity of the resonance.
The relation between the partial-wave decay amplitude $G(\ell)$ and the decay width 
$\Gamma (R \to \Delta \pi)$ is given by
%EQUATION>>>
\begin{align}
\label{eq:DA2_Sum}
\Gamma (R \to \Delta \pi) = \sum_\ell |G(\ell)|^2 .
\end{align}
%EQUATION<<<

We write the general form of the interaction Lagrangians and the relevant decay
widths and partial-wave decay amplitudes below.

\subsubsection{Spin-$1/2$ resonance}

The interaction Lagrangian for spin-$1/2$ resonance can be written as
%EQUATION>>>
\begin{align}
\mathcal{L}^{(1/2\pm)}_{R\Delta\pi} = 
\frac{h_1}{M_\pi} \bar{\Delta}^\mu \Gamma^{(\mp)} \partial_\mu \pi R + \mbox{H.c.},
\end{align}
%EQUATION<<<
which leads to
%EQUATION>>>
\begin{align}
\Gamma({\textstyle\frac12}^\pm \to \Delta\pi) =
\frac{h_1^2}{6\pi} \frac{q_\pi^3 M_R}{M_\pi^2 M_\Delta^2} (E_\Delta \pm M_\Delta).
\end{align}
%EQUATION<<<

The final state is in the relative $p$ and $d$ waves in the decays of positive and
negative resonances, respectively.
Threrefore, we have
%EQUATION>>>
\begin{align}
G(1) = - \frac{1}{\sqrt{6\pi}} \frac{q_\pi}{M_\Delta}
\sqrt{q_\pi M_R} \sqrt{E_\Delta+M_\Delta} \frac{h_1}{M_\pi}, 
\end{align}
%EQUATION<<<
for a $j^P = \textstyle\frac{1}{2}^+ $ resonance, and
%EQUATION>>>
\begin{align}
G(2) = \frac{1}{\sqrt{6\pi}} \frac{q_\pi}{M_\Delta}
\sqrt{q_\pi M_R} \sqrt{E_\Delta-M_\Delta} \frac{h_1}{M_\pi},
\end{align}
%EQUATION<<<
for a $j^P = \textstyle\frac{1}{2}^-$ resonance.

\subsubsection{Spin-$3/2$ resonance}

The interaction Lagrangian for spin-$3/2$ resonance can be written as
%EQUATION>>>
\begin{align}
\mathcal{L}^{(3/2\pm)}_{R\Delta\pi} =&
\left[
\frac{h_1}{M_\pi} \bar\Delta^\mu \Gamma_\nu^{(\pm)}
+ \frac{ih_2}{M_\pi^2} \bar\Delta_\nu \Gamma^{(\pm)} \partial^\mu
\right]
\cr   & \times \partial^\nu \pi R_\mu 
+ \mbox{H.c.},
\end{align}
%EQUATION<<<
which leads to
%EQUATION>>>
\begin{align}
\Gamma({\textstyle\frac32}^\pm \to \Delta\pi) =&
\frac{1}{36\pi} \frac{q_\pi}{M_R M_\Delta^2} (E_\Delta \mp M_\Delta) 
\cr   & \hspace{-6em}
\times \biggl[ \frac{h_1^2}{M_\pi^2} (M_R \pm M_\Delta)^2
(2E_\Delta^2 \mp 2 E_\Delta M_\Delta + 5 M_\Delta^2)
\cr   & \hspace{-6em}
\mp 2 \frac{h_1 h_2}{M_\pi^3} M_R q_\pi^2 (M_R \pm M_\Delta) (2E_\Delta \mp M_\Delta)
\cr   & \hspace{-6em}
+ 2 \frac{h_2^2}{M_\pi^4} M_R^2 q_\pi^4 \biggr].
\end{align}
%EQUATION<<<

The final state is in the relative $p$ and $f$ waves in the decay of a positive parity
resonance and is in the relative $s$ and $d$ waves in the decay of a negative resonance.
The partial-wave decay amplitude can be expressed in terms of the coupling constants as
%EQUATION>>>
\begin{align}
G(1) =& G_{11}^{(3/2)}\frac{h_1}{M_\pi} + G_{12}^{(3/2)}\frac{h_2}{M_\pi^2},   \cr
G(3) =& G_{31}^{(3/2)}\frac{h_1}{M_\pi} + G_{32}^{(3/2)}\frac{h_2}{M_\pi^2},
\end{align}
%EQUATION<<<
for a $j^P = \textstyle\frac{3}{2}^+ $ resonance, where
%EQUATION>>>
\begin{align}
G_{11}^{(3/2)} =& \frac{\sqrt{5}}{30\sqrt\pi} 
\frac{1}{M_\Delta} \sqrt{\frac{q_\pi}{M_R} } \sqrt{E_\Delta-M_\Delta}
\cr   &
\times (M_R+M_\Delta) (E_\Delta+4M_\Delta),
\cr
G_{12}^{(3/2)} =& - \frac{\sqrt{5}}{30\sqrt\pi} 
\frac{q_\pi^2 \sqrt{q_\pi M_R}}{M_\Delta} \sqrt{E_\Delta-M_\Delta},
\cr
G_{31}^{(3/2)} =& - \frac{\sqrt{5}}{10\sqrt\pi} 
\frac{1}{M_\Delta} \sqrt{\frac{q_\pi}{M_R} } \sqrt{E_\Delta-M_\Delta}
\cr   &
\times (M_R+M_\Delta) (E_\Delta-M_\Delta),
\cr
G_{32}^{(3/2)} =& \frac{\sqrt{5}}{10\sqrt\pi} 
\frac{q_\pi^2 \sqrt{q_\pi M_R}}{M_\Delta} \sqrt{E_\Delta-M_\Delta},
\end{align}
%EQUATION<<<
and
%EQUATION>>>
\begin{align}
G(0) =& G_{01}^{(3/2)}\frac{h_1}{M_\pi} + G_{02}^{(3/2)}\frac{h_2}{M_\pi^2},   \cr
G(2) =& G_{21}^{(3/2)}\frac{h_1}{M_\pi} + G_{22}^{(3/2)}\frac{h_2}{M_\pi^2},
\end{align}
%EQUATION<<<
for a $j^P = \textstyle\frac{3}{2}^- $ resonance, where
%EQUATION>>>
\begin{align}
G_{01}^{(3/2)} =& \frac{1}{6\sqrt\pi}
\frac{1}{M_\Delta} \sqrt{\frac{q_\pi}{M_R}} \sqrt{E_\Delta+M_\Delta}
\cr   &
\times (M_R-M_\Delta) (E_\Delta+2M_\Delta),
\cr
G_{02}^{(3/2)} =& \frac{1}{6\sqrt\pi} 
\frac{q_\pi^2 \sqrt{q_\pi M_R}}{M_\Delta} \sqrt{E_\Delta+M_\Delta},
\cr
G_{21}^{(3/2)} =& - \frac{1}{6\sqrt\pi} 
\frac{1}{M_\Delta} \sqrt{\frac{q_\pi}{M_R}} \sqrt{E_\Delta+M_\Delta}
\cr   &
\times (M_R-M_\Delta) (E_\Delta-M_\Delta),
\cr
G_{22}^{(3/2)} =& - \frac{1}{6\sqrt\pi}
\frac{q_\pi^2 \sqrt{q_\pi M_R}}{M_\Delta} \sqrt{E_\Delta+M_\Delta}.
\end{align}
%EQUATION<<<

\subsubsection{Spin-$5/2$ resonance}

The interaction Lagrangian for spin-$5/2$ resonance can be written as
%EQUATION>>>
\begin{align}
\mathcal{L}^{(5/2\pm)}_{R\Delta\pi} =&
\left[
\frac{ih_1}{M_\pi^2} \bar\Delta^\mu \Gamma_\nu^{(\mp)}
- \frac{h_2}{M_\pi^3} \bar\Delta_\nu \Gamma^{(\mp)} \partial^\mu
\right]
\cr   & \times \partial^\nu \partial^\alpha \pi R_{\mu\alpha}
+ \mbox{H.c.},
\end{align}
%EQUATION<<<
which leads to
%EQUATION>>>
\begin{align}
\Gamma({\textstyle\frac52}^\pm \to \Delta\pi) =&
\frac{1}{180\pi} \frac{q_\pi^3}{M_R M_\Delta^2} (E_\Delta \pm M_\Delta) 
\cr   & \hspace{-6em}
\times \biggl[ \frac{h_1^2}{M_\pi^4} (M_R \mp M_\Delta)^2
(4E_\Delta^2 \pm 4 E_\Delta M_\Delta + 7 M_\Delta^2)
\cr   & \hspace{-6em}
\pm 4 \frac{h_1 h_2}{M_\pi^5} M_R q_\pi^2 (M_R \mp M_\Delta) (2E_\Delta \pm M_\Delta)
\cr   & \hspace{-6em}
+ 4 \frac{h_2^2}{M_\pi^6} M_R^2 q_\pi^4 \biggr].
\end{align}
%EQUATION<<<

The final state is in the relative $p$ and $f$ waves in the decay of a positive parity
resonance and is in the relative $d$ and $g$ waves in the decay of a negative resonance.
The partial-wave decay amplitude can be expressed as
%EQUATION>>>
\begin{align}
G(1) =& G_{11}^{(5/2)}\frac{h_1}{M_\pi^2} + G_{12}^{(5/2)}\frac{h_2}{M_\pi^3},   \cr
G(3) =& G_{31}^{(5/2)}\frac{h_1}{M_\pi^2} + G_{32}^{(5/2)}\frac{h_2}{M_\pi^3},
\end{align}
%EQUATION<<<
for a $j^P = \textstyle\frac{5}{2}^+ $ resonance, where
%EQUATION>>>
\begin{align}
G_{11}^{(5/2)} =& \frac{\sqrt{3}}{30\sqrt\pi}
\frac{q_\pi}{M_\Delta} \sqrt{\frac{q_\pi}{M_R} } \sqrt{E_\Delta+M_\Delta}
\cr   &
\times (M_R-M_\Delta) (2E_\Delta+3M_\Delta),
\cr
G_{12}^{(5/2)} =& \frac{\sqrt{3}}{15\sqrt\pi}
\frac{q_\pi^3 \sqrt{q_\pi M_R}}{M_\Delta} \sqrt{E_\Delta+M_\Delta},
\cr
G_{31}^{(5/2)} =& - \frac{\sqrt{2}}{15\sqrt\pi} 
\frac{q_\pi}{M_\Delta} \sqrt{\frac{q_\pi}{M_R} } \sqrt{E_\Delta+M_\Delta}
\cr   &
\times (M_R-M_\Delta) (E_\Delta-M_\Delta),
\cr
G_{32}^{(5/2)} =& - \frac{\sqrt{2}}{15\sqrt\pi} 
\frac{q_\pi^3 \sqrt{q_\pi M_R}}{M_\Delta} \sqrt{E_\Delta+M_\Delta},
\end{align}
%EQUATION<<<
and
%EQUATION>>>
\begin{align}
G(2) =& G_{21}^{(5/2)}\frac{h_1}{M_\pi^2} + G_{22}^{(5/2)}\frac{h_2}{M_\pi^3},   \cr
G(4) =& G_{41}^{(5/2)}\frac{h_1}{M_\pi^2} + G_{42}^{(5/2)}\frac{h_2}{M_\pi^3},
\end{align}
%EQUATION<<<
for a $j^P = \textstyle\frac{5}{2}^- $ resonance, where
%EQUATION>>>
\begin{align}
G_{21}^{(5/2)} =& \frac{\sqrt{35}}{210\sqrt\pi}
\frac{q_\pi}{M_\Delta} \sqrt{\frac{q_\pi}{M_R} } \sqrt{E_\Delta-M_\Delta}
\cr   &
\times (M_R+M_\Delta) (2E_\Delta+5M_\Delta),
\cr
G_{22}^{(5/2)} =& - \frac{\sqrt{35}}{105\sqrt\pi}
\frac{q_\pi^3 \sqrt{q_\pi M_R}}{M_\Delta} \sqrt{E_\Delta-M_\Delta},
\cr
G_{41}^{(5/2)} =& - \frac{\sqrt{210}}{105\sqrt\pi} 
\frac{q_\pi}{M_\Delta} \sqrt{\frac{q_\pi}{M_R} } \sqrt{E_\Delta-M_\Delta}
\cr   &
\times (M_R+M_\Delta) (E_\Delta-M_\Delta),
\cr
G_{42}^{(5/2)} =& \frac{\sqrt{210}}{105\sqrt\pi} 
\frac{q_\pi^3 \sqrt{q_\pi M_R}}{M_\Delta} \sqrt{E_\Delta-M_\Delta}.
\end{align}
%EQUATION<<<

\subsubsection{Spin-$7/2$ resonance}

The interaction Lagrangian for spin-$7/2$ resonance can be written as
%EQUATION>>>
\begin{align}
\mathcal{L}^{(7/2\pm)}_{R\Delta\pi} =&
\left[
-\frac{h_1}{M_\pi^3} \bar\Delta^\mu \Gamma_\nu^{(\pm)}
- \frac{ih_2}{M_\pi^4} \bar\Delta_\nu \Gamma^{(\pm)} \partial^\mu
\right]
\cr   & \times \partial^\nu \partial^\alpha \partial^\beta \pi R_{\mu\alpha\beta} 
+ \mbox{H.c.},
\end{align}
%EQUATION<<<
which leads to
%EQUATION>>>
\begin{align}
\Gamma({\textstyle\frac72}^\pm \to \Delta\pi) =&
\frac{1}{210\pi} \frac{q_\pi^5}{M_R M_\Delta^2} (E_\Delta \mp M_\Delta) 
\cr   & \hspace{-6em}
\times \biggl[ \frac{h_1^2}{M_\pi^6} (M_R \pm M_\Delta)^2
(2E_\Delta^2 \mp 2 E_\Delta M_\Delta + 3 M_\Delta^2)
\cr   & \hspace{-6em}
\mp 2 \frac{h_1 h_2}{M_\pi^7} M_R q_\pi^2 (M_R \pm M_\Delta) (2E_\Delta \mp M_\Delta)
\cr   & \hspace{-6em}
+ 2 \frac{h_2^2}{M_\pi^8} M_R^2 q_\pi^4 \biggr].
\end{align}
%EQUATION<<<

The final state is in the relative $f$ and $h$ waves in the decay of a positive parity
resonance and is in the relative $d$ and $g$ waves in the decay of a negative resonance.
The partial-wave decay amplitude can be expressed as
%EQUATION>>>
\begin{align}
G(3) =& G_{31}^{(7/2)}\frac{h_1}{M_\pi^3} + G_{32}^{(7/2)}\frac{h_2}{M_\pi^4} ,  \cr
G(5) =& G_{51}^{(7/2)}\frac{h_1}{M_\pi^3} + G_{52}^{(7/2)}\frac{h_2}{M_\pi^4},
\end{align}
%EQUATION<<<
for a $j^P = \textstyle\frac{7}{2}^+ $ resonance, where
%EQUATION>>>
\begin{align}
G_{31}^{(7/2)} =& \frac{\sqrt{70}}{210\sqrt\pi}
\frac{q_\pi^2}{M_\Delta} \sqrt{\frac{q_\pi}{M_R} } \sqrt{E_\Delta-M_\Delta}
\cr   &
\times (M_R+M_\Delta) (E_\Delta+2M_\Delta),
\cr
G_{32}^{(7/2)} =& - \frac{\sqrt{70}}{210\sqrt\pi}
\frac{q_\pi^4 \sqrt{q_\pi M_R}}{M_\Delta} \sqrt{E_\Delta-M_\Delta},
\cr
G_{51}^{(7/2)} =& - \frac{\sqrt{14}}{42\sqrt\pi} 
\frac{q_\pi^2}{M_\Delta} \sqrt{\frac{q_\pi}{M_R} } \sqrt{E_\Delta-M_\Delta}
\cr   &
\times (M_R+M_\Delta) (E_\Delta-M_\Delta),
\cr
G_{52}^{(7/2)} =& \frac{\sqrt{14}}{42\sqrt\pi} 
\frac{q_\pi^4 \sqrt{q_\pi M_R}}{M_\Delta} \sqrt{E_\Delta-M_\Delta},
\end{align}
%EQUATION<<<
and
%EQUATION>>>
\begin{align}
G(2) =& G_{21}^{(7/2)}\frac{h_1}{M_\pi^3} + G_{22}^{(7/2)}\frac{h_2}{M_\pi^4},   \cr
G(4) =& G_{41}^{(7/2)}\frac{h_1}{M_\pi^3} + G_{42}^{(7/2)}\frac{h_2}{M_\pi^4},
\end{align}
%EQUATION<<<
for a $j^P = \textstyle\frac{7}{2}^- $ resonance, where
%EQUATION>>>
\begin{align}
G_{21}^{(7/2)} =& \frac{\sqrt{210}}{210\sqrt\pi}
\frac{q_\pi^2}{M_\Delta} \sqrt{\frac{q_\pi}{M_R} } \sqrt{E_\Delta+M_\Delta}
\cr   &
\times (M_R-M_\Delta) (3E_\Delta+4M_\Delta),
\cr
G_{22}^{(7/2)} =& \frac{\sqrt{30}}{70\sqrt\pi}
\frac{q_\pi^4 \sqrt{q_\pi M_R}}{M_\Delta} \sqrt{E_\Delta+M_\Delta},
\cr
G_{41}^{(7/2)} =& - \frac{\sqrt{6}}{42\sqrt\pi} 
\frac{q_\pi^2}{M_\Delta} \sqrt{\frac{q_\pi}{M_R} } \sqrt{E_\Delta+M_\Delta}
\cr   &
\times (M_R-M_\Delta) (E_\Delta-M_\Delta),
\cr
G_{42}^{(7/2)} =& - \frac{\sqrt{6}}{42\sqrt\pi} 
\frac{q_\pi^4 \sqrt{q_\pi M_R}}{M_\Delta} \sqrt{E_\Delta+M_\Delta}.
\end{align}
%EQUATION<<<

\subsubsection{Spin-$9/2$ resonance}

The interaction Lagrangian for spin-$9/2$ resonance can be written as
%EQUATION>>>
\begin{align}
\mathcal{L}^{(9/2\pm)}_{R\Delta\pi} =&
\left[
-\frac{ih_1}{M_\pi^4} \bar\Delta^\mu \Gamma_\nu^{(\mp)}
+ \frac{h_2}{M_\pi^5} \bar\Delta_\nu \Gamma^{(\mp)} \partial^\mu
\right]
\cr   & \times
\partial^\nu \partial^\alpha \partial^\beta \partial^\rho \pi R_{\mu\alpha\beta\rho}
+ \mbox{H.c.},
\end{align}
%EQUATION<<<
which leads to
%EQUATION>>>
\begin{align}
\Gamma({\textstyle\frac92}^\pm \to \Delta\pi) =&
\frac{1}{1890\pi} \frac{q_\pi^7}{M_R M_\Delta^2} (E_\Delta \pm M_\Delta)
\cr   & \hspace{-6em}
\times \biggl[ \frac{h_1^2}{M_\pi^8} (M_R \mp M_\Delta)^2
(8E_\Delta^2 \pm 8 E_\Delta M_\Delta + 11 M_\Delta^2)
\cr   & \hspace{-6em}
\pm 8 \frac{h_1 h_2}{M_\pi^9} M_R q_\pi^2 (M_R \mp M_\Delta) (2E_\Delta \pm M_\Delta)
\cr   & \hspace{-6em}
+ 8 \frac{h_2^2}{M_\pi^{10}} M_R^2 q_\pi^4 \biggr].
\end{align}
%EQUATION<<<

The final state is in the relative $f$ and $h$ waves in the decay of a positive parity
resonance and is in the relative $g$ and $i$ waves in the decay of a negative resonance.
The partial-wave decay amplitude can be expressed as
%EQUATION>>>
\begin{align}
G(3) =& G_{31}^{(9/2)}\frac{h_1}{M_\pi^4} + G_{32}^{(9/2)}\frac{h_2}{M_\pi^5},   \cr
G(5) =& G_{51}^{(9/2)}\frac{h_1}{M_\pi^4} + G_{52}^{(9/2)}\frac{h_2}{M_\pi^5},
\end{align}
%EQUATION<<<
for a $j^P = \textstyle\frac{9}{2}^+ $ resonance, where
%EQUATION>>>
\begin{align}
G_{31}^{(9/2)} =& \frac{\sqrt{70}}{630\sqrt\pi}
\frac{q_\pi^3}{M_\Delta} \sqrt{\frac{q_\pi}{M_R} } \sqrt{E_\Delta+M_\Delta}
\cr   &
\times (M_R-M_\Delta) (4E_\Delta+5M_\Delta),
\cr
G_{32}^{(9/2)} =& \frac{2\sqrt{70}}{315\sqrt\pi}
\frac{q_\pi^5 \sqrt{q_\pi M_R}}{M_\Delta} \sqrt{E_\Delta+M_\Delta},
\cr
G_{51}^{(9/2)} =& - \frac{2\sqrt{35}}{315\sqrt\pi} 
\frac{q_\pi^3}{M_\Delta} \sqrt{\frac{q_\pi}{M_R} } \sqrt{E_\Delta+M_\Delta}
\cr   &
\times (M_R-M_\Delta) (E_\Delta-M_\Delta),
\cr
G_{52}^{(9/2)} =& - \frac{2\sqrt{35}}{315\sqrt\pi} 
\frac{q_\pi^5 \sqrt{q_\pi M_R}}{M_\Delta} \sqrt{E_\Delta+M_\Delta},
\end{align}
%EQUATION<<<
and
%EQUATION>>>
\begin{align}
G(4) =& G_{41}^{(9/2)}\frac{h_1}{M_\pi^4} + G_{42}^{(9/2)}\frac{h_2}{M_\pi^5},   \cr
G(6) =& G_{61}^{(9/2)}\frac{h_1}{M_\pi^4} + G_{62}^{(9/2)}\frac{h_2}{M_\pi^5},
\end{align}
%EQUATION<<<
for a $j^P = \textstyle\frac{9}{2}^- $ resonance, where
%EQUATION>>>
\begin{align}
G_{41}^{(9/2)} =& \frac{\sqrt{2310}}{6930\sqrt\pi}
\frac{q_\pi^3}{M_\Delta} \sqrt{\frac{q_\pi}{M_R} } \sqrt{E_\Delta-M_\Delta}
\cr   &
\times (M_R+M_\Delta) (4E_\Delta+7M_\Delta),
\cr
G_{42}^{(9/2)} =& - \frac{2\sqrt{2310}}{3465\sqrt\pi}
\frac{q_\pi^5 \sqrt{q_\pi M_R}}{M_\Delta} \sqrt{E_\Delta-M_\Delta},
\cr
G_{61}^{(9/2)} =& - \frac{2\sqrt{1155}}{1155\sqrt\pi} 
\frac{q_\pi^3}{M_\Delta} \sqrt{\frac{q_\pi}{M_R} } \sqrt{E_\Delta-M_\Delta}
\cr   &
\times (M_R+M_\Delta) (E_\Delta-M_\Delta),
\cr
G_{62}^{(9/2)} =& \frac{2\sqrt{1155}}{1155\sqrt\pi} 
\frac{q_\pi^5 \sqrt{q_\pi M_R}}{M_\Delta} \sqrt{E_\Delta-M_\Delta}.
\end{align}
%EQUATION<<<

\subsection{Vector meson couplings}

For the decay of $R \to \Delta V$, there are 3 independent couplings for $1/2^P \to
1^- + 3/2^+$, 5 independent terms for $3/2^P \to 1^- + 3/2^+ $, and 6 independent
terms for $j^P \to 1^- + 3/2^+$ with $j \ge 5/2$, as indicated in Table~\ref{TAB02}.

The helicity amplitudes for the decay of $R \to \Delta V$ are related to the decay
width as
%EQUATION>>>
\begin{align}
\Gamma(R \to \Delta V) =& \frac{q_V^2}{\pi} \frac{2M_N}{(2j+1) M_R}
\biggl\{ \left| A_{1/2} \right|^2 + \left| S_{1/2} \right|^2
\cr   &
\hspace{-3.5em}
+ \left| D_{1/2} \right|^2 + \left| A_{3/2} \right|^2 
+ \left| S_{3/2} \right|^2 + \left| D_{5/2} \right|^2
\biggr\},
\end{align}
%EQUATION<<<
where $q_V = |{\bf q}_V|$ is the magnitude of the three-momentum of the vector
meson in the rest frame of the resonance
%EQUATION>>>
\begin{align}
q_V = \frac{1}{2M_R} 
\sqrt{[M_R^2 - (M_\Delta+M_V)^2][M_R^2 - (M_\Delta-M_V)^2]}.
\end{align}
%EQUATION<<<
The helicity amplitudes are then obtained as
%EQUATION>>>
\begin{align}
A_\lambda^{(j\pm)} =& \frac{1}{\sqrt{8M_N M_R q_V}} \frac{2j+1}{4\pi}
\cr   & \times
\int d \cos\theta\, d\phi\, e^{-i(m-\lambda)\phi} d^j_{\lambda m}(\theta) 
\cr   & \times
\langle \textbf{q}_V, \lambda_V,\lambda_\Delta \mid -i \mathcal{M}_{R \to \Delta V} \mid
j m \rangle,
\label{eq:HA-2}
\end{align}
%EQUATION<<<
where $\mathcal{M}_{R \to \Delta V}$ is the transition amplitude.
The lower index stands for the helicty of the final $\Delta V$ state, i.e.,
$\lambda = \lambda_V - \lambda_\Delta$, where we use an abbraviation as follows
%EQUATION>>>
\begin{align}
& A_{1/2} : 1  - \textstyle\frac12,\,\,\,
  S_{1/2} : 0  +  \textstyle\frac12,\,\,\,
  D_{1/2} : -1 + \textstyle\frac32,
\cr
& A_{3/2} : 1 + \textstyle\frac12,\,\,\,
  S_{3/2} : 0 + \textstyle\frac32,\,\,\,
  D_{5/2} : 1 + \textstyle\frac32.
\end{align}
%EQUATION<<<

The details of the interaction Lagrangians and the relevant helicity amplitudes are
given below.

%------------------------------>>> widetext
\begin{widetext}

\subsubsection{Effective Lagrangians}

The effective Lagrangian for the $R \Delta V$ interaction can be constructed by using
the field strength tensor $V_{\mu\nu} = \partial_\mu V_\nu - \partial_\nu V_\mu$ as
follows
\begin{align}
\mathcal{L}_{R \Delta V}^{(1/2\pm)} =&
\left[
\frac{-ig_1}{2M_N} \bar\Delta_\mu \Gamma_\nu^{(\pm)}
- \frac{g_2}{(2M_N)^2} \partial_\nu \bar\Delta_\mu \Gamma^{(\pm)}
+ \frac{g_3}{(2M_N)^2} \bar\Delta_\mu \Gamma^{(\pm)} \partial_\nu
\right]
V^{\mu\nu} R + \mbox{H.c.},
\cr
\mathcal{L}_{R \Delta V}^{(3/2\pm)} = &
\left[
\frac{g_1}{(2M_N)^2} \bar\Delta_\mu \Gamma_\nu^{(\mp)}
- \frac{ig_2}{(2M_N)^3} \partial_\nu \bar\Delta_\mu \Gamma^{(\mp)}
+ \frac{ig_3}{(2M_N)^3} \bar\Delta_\mu \Gamma^{(\mp)} \partial_\nu
\right]
\partial^\alpha V^{\mu\nu} R_\alpha
\cr &
- \left[
\frac{g_4}{(2M_N)^2} \partial_\nu \bar\Delta^\alpha \Gamma_\mu^{(\mp)}
- \frac{g_5}{(2M_N)^2} \bar\Delta^\alpha \Gamma_\mu^{(\mp)}
\partial_\nu
\right]
V^{\mu\nu} R_\alpha + \mbox{H.c.},
\cr
\mathcal{L}_{R \Delta V}^{(5/2\pm)} =&
\left[
\frac{ig_1}{(2M_N)^3} \bar\Delta_\mu \Gamma_\nu^{(\pm)}
+ \frac{g_2}{(2M_N)^4} \partial_\nu \bar\Delta_\mu \Gamma^{(\pm)}
- \frac{g_3}{(2M_N)^4} \bar\Delta_\mu \Gamma^{(\pm)} \partial_\nu
\right]
\partial^\alpha \partial^\beta
V^{\mu\nu} R_{\alpha\beta}
\cr &
- \left[
\frac{ig_4}{(2M_N)^3} \partial_\nu \bar\Delta^\alpha \Gamma_\mu^{(\pm)}
- \frac{ig_5}{(2M_N)^3} \bar\Delta^\alpha \Gamma_\mu^{(\pm)} \partial_\nu
\right]
\partial^\beta V^{\mu\nu} R_{\alpha\beta}
- \frac{ig_6}{2M_N} \bar\Delta^\alpha \Gamma_\mu^{(\pm)}
V^{\mu\beta}
R_{\alpha\beta} + \mbox{H.c.},
\cr
\mathcal{L}_{R \Delta V}^{(7/2\pm)} =&
\left[
- \frac{g_1}{(2M_N)^4} \bar\Delta_\mu \Gamma_\nu^{(\mp)}
+ \frac{ig_2}{(2M_N)^5} \partial_\nu \bar\Delta_\mu \Gamma^{(\mp)}
- \frac{ig_3}{(2M_N)^5} \bar\Delta_\mu \Gamma^{(\mp)} \partial_\nu
\right]
\partial^\alpha \partial^\beta \partial^\rho
V^{\mu\nu} R_{\alpha \beta \rho}
\cr &
+ \left[
\frac{g_4}{(2M_N)^4} \partial_\nu \bar\Delta^\alpha \Gamma_\mu^{(\mp)}
- \frac{g_5}{(2M_N)^4} \bar\Delta^\alpha \Gamma_\mu^{(\mp)} \partial_\nu
\right]
\partial^\beta \partial^\rho V^{\mu\nu}
R_{\alpha \beta \rho}
+ \frac{g_6}{(2M_N)^2} \bar\Delta^\alpha \Gamma_\mu^{(\mp)}
\partial^\rho V^{\mu\beta}
R_{\alpha \beta \rho} + \mbox{H.c.},
\cr
\mathcal{L}_{R \Delta V}^{(9/2\pm)} =&
\left[
- \frac{ig_1}{(2M_N)^5} \bar\Delta_\mu \Gamma_\nu^{(\pm)}
- \frac{g_2}{(2M_N)^6} \partial_\nu \bar\Delta_\mu \Gamma^{(\pm)}
+ \frac{g_3}{(2M_N)^6} \bar\Delta_\mu \Gamma^{(\pm)} \partial_\nu
\right]
\partial^{\alpha} \partial^\beta
\partial^\rho \partial^\delta
V^{\mu\nu}
R_{\alpha \beta \rho \delta}
\\ &
+ \left[
\frac{ig_4}{(2M_N)^5} \partial_\nu \bar\Delta^\alpha \Gamma_\mu^{(\pm)}
- \frac{ig_5}{(2M_N)^5} \bar\Delta^\alpha \Gamma_\mu^{(\pm)} \partial_\nu
\right]
\partial^\beta \partial^\rho \partial^\delta V^{\mu\nu}
R_{\alpha \beta \rho \delta}
+ \frac{ig_6}{(2M_N)^3} \bar\Delta^\alpha \Gamma_\mu^{(\pm)}
\partial^\rho \partial^\delta V^{\mu \beta}
R_{\alpha \beta \rho \delta} + \mbox{H.c.}.
\nonumber
\end{align}
In the case of $j \geq 7/2$, it is obtained by taking more derivatives on the vector
field.

\subsubsection{Helicity amplitudes}

The helicity amplitudes derived from Eq.~(\ref{eq:HA-2}) are given as follows

\textbullet\ spin-$1/2$
\begin{subequations}
%EQUATION>>>
\begin{align}
A_{1/2}^{(1/2\pm)} =& - \frac{1}{4\sqrt3}
\frac{\sqrt{E_\Delta \mp M_\Delta}}{M_N \sqrt{q_V M_N}}
\left[
  g_1^{} \frac{M_R(M_R \pm M_\Delta) - M_V^2}{M_\Delta}
+ g_2^{} \frac{M_R^2 - M_\Delta^2 - M_V^2}{4M_N} - g_3^{} \frac{M_V^2}{2 M_N}
\right],
\cr
S_{1/2}^{(1/2\pm)} =& \frac{1}{2\sqrt6}
\frac{\sqrt{E_\Delta \mp M_\Delta}}{M_N M_\Delta \sqrt{q_V M_N}}
\biggl[
 g_1^{} M_\Delta M_V +
 g_2^{} \frac{M_R^2(M_R^2-4E_\Delta^2+4M_\Delta^2)+(M_\Delta^2+M_V^2)(M_\Delta^2+M_V^2-2M_R^2)}
{8 M_N M_V}
\cr &\hspace{10em}
- g_3^{} \frac{M_V(M_R^2 -M_\Delta^2 - M_V^2)}{4M_N}
\biggr],
\cr
D_{1/2}^{(1/2\pm)} =& \pm \frac{1}{4}
\frac{\sqrt{E_\Delta \mp M_\Delta}}{M_N \sqrt{q_V M_N}}
\left[
g_1^{} (M_R \pm M_\Delta) \mp g_2^{} \frac{M_R^2 -M_\Delta^2 - M_V^2}{4M_N}
\pm g_3^{} \frac{M_V^2}{2M_N}
\right].
%EQUATION<<<
\end{align}
\end{subequations}

\textbullet\ spin-$3/2$
\begin{subequations}
%EQUATION>>>
\begin{align}
A_{1/2}^{(3/2\pm)} =& - \frac{1}{12\sqrt2}
\frac{\sqrt{E_\Delta \mp M_\Delta}}{M_N^2 \sqrt{q_V M_N}}
\biggl\{ (E_\Delta \pm M_\Delta)
\left[
  g_1^{} \frac{M_R(M_R \mp M_\Delta) - M_V^2}{M_\Delta}
+ g_2^{} \frac{M_R^2 - M_\Delta^2 - M_V^2}{4 M_N}
- g_3^{} \frac{M_V^2}{2 M_N}
\right]
\cr  & \hspace{10.0em}
- \frac{E_\Delta}{M_\Delta}
\left[
  g_4^{} (M_R^2 - M_\Delta^2 - M_V^2)
- g_5^{} (2 M_V^2) \right] \biggr\},
\cr
S_{1/2}^{(3/2\pm)} =& \frac{1}{12}
\frac{\sqrt{E_\Delta \mp M_\Delta}}{M_N^2 M_\Delta \sqrt{q_V M_N}}
\biggl\{ (E_\Delta \pm M_\Delta)
\biggl[
g_1^{} M_\Delta M_V
\cr & \hspace{9.0em}
+ g_2^{}
\frac{M_R^2(M_R^2-4E_\Delta^2+4M_\Delta^2)+(M_\Delta^2+M_V^2)(M_\Delta^2+M_V^2-2M_R^2)}
{8 M_N M_V}
\cr & \hspace{9.0em}
- g_3^{} \frac{M_V (M_R^2 - M_\Delta^2 - M_V^2)}{4M_N}
\biggr]
- \frac{(2 E_\Delta \pm M_\Delta) M_V}{2} [g_4^{} M_\Delta \pm g_5^{} (M_R \pm M_\Delta)]
\biggr\},
\cr
D_{1/2}^{(3/2\pm)} =& \mp \frac{1}{4\sqrt6}
\frac{\sqrt{E_\Delta \mp M_\Delta}}{M_N^2 \sqrt{q_V M_N}}
\biggl\{ (E_\Delta \pm M_\Delta)
\left[
    g_1^{} (M_R \mp M_\Delta)
\pm g_2^{} \frac{M_R^2 -M_\Delta^2 - M_V^2}{4M_N}
\mp g_3^{} \frac{M_V^2}{2M_N}
\right]
\cr & \hspace{9.5em}
- \left[
g_4^{} \frac{M_R^2 - M_\Delta^2 - M_V^2}{2}
- g_5^{} M_V^2
\right]
\biggr\},
\cr
A_{3/2}^{(3/2\pm)} =& \frac{1}{8\sqrt6}
\frac{\sqrt{E_\Delta \mp M_\Delta}}{M_N^2 \sqrt{q_V M_N}}
\left[ g_4^{} (M_R^2 - M_\Delta^2 - M_V^2) - g_5^{} (2 M_V^2) \right],
\cr
S_{3/2}^{(3/2\pm)} =& -\frac{1}{8}
\frac{\sqrt{E_\Delta \mp M_\Delta}M_V}{M_N^2 \sqrt{q_V M_N}}
\left[ g_4^{} M_\Delta \pm g_5^{} (M_R \pm M_\Delta) \right],
\end{align}
%EQUATION<<<
\end{subequations}
where, for the $g_1$, $g_2$, and $g_3$ couplings, which are contained
only for $A_{1/2}^{(3/2\pm)}$, $S_{1/2}^{(3/2\pm)}$, and $D_{1/2}^{(3/2\pm)}$, the following
relations are satisfied
%EQUATION>>>
\begin{align}
A_{1/2}^{(3/2\pm)} = \frac{1}{\sqrt{6}} \frac{q_V}{M_N} A_{1/2}^{(1/2\mp)},\,\,\,
S_{1/2}^{(3/2\pm)} = \frac{1}{\sqrt{6}} \frac{q_V}{M_N} S_{1/2}^{(1/2\mp)},\,\,\,
D_{1/2}^{(3/2\pm)} = \frac{1}{\sqrt{6}} \frac{q_V}{M_N} D_{1/2}^{(1/2\mp)}.
\end{align}
%EQUATION<<<

\textbullet\ spin-$5/2$
\begin{subequations}
%EQUATION>>>
\begin{align}
A_{1/2}^{(5/2\pm)} =& - \frac{1}{8\sqrt30}
\frac{\sqrt{E_\Delta \mp M_\Delta}}{M_N^3 \sqrt{q_V M_N}}
\biggl\{ q_V^2
\left[
  g_1^{} \frac{M_R(M_R \pm  M_\Delta) - M_V^2}{M_\Delta}
+ g_2^{} \frac{M_R^2 - M_\Delta^2 - M_V^2}{4 M_N}
- g_3^{} \frac{M_V^2}{2 M_N}
\right]
\cr & \hspace{10.0em}
- (E_\Delta \pm M_\Delta) \frac{E_\Delta}{M_\Delta}
\left[ g_4^{} (M_R^2 - M_\Delta^2 - M_V^2) - g_5^{} (2 M_V^2) \right]
\cr &\hspace{10.0em}
+ g_6^{} \frac{2M_N^2}{M_\Delta}
[ (2E_\Delta \pm M_\Delta)(E_\Delta \pm M_\Delta)+(2E_\Delta \mp M_\Delta)(E_\Delta-M_R) ]
\biggr\},
\cr
S_{1/2}^{(5/2\pm)} =& \frac{1}{8\sqrt{15}}
\frac{\sqrt{E_\Delta \mp M_\Delta}}{M_N^3 M_\Delta \sqrt{q_V M_N}}
\biggl\{
q_V^2
\biggl[
g_1^{} M_\Delta M_V
\cr &
+ g_2^{}
\frac{M_R^2(M_R^2-4E_\Delta^2+4M_\Delta^2)+(M_\Delta^2+M_V^2)(M_\Delta^2+M_V^2-2M_R^2)}
{8 M_N M_V}
- g_3^{} \frac{M_V (M_R^2 - M_\Delta^2 - M_V^2)}{4M_N}
\biggr]
\cr &
- (E_\Delta \pm M_\Delta) \frac{(2 E_\Delta \mp M_\Delta) M_V}{2}
[g_4^{} M_\Delta \mp g_5^{} (M_R \mp M_\Delta)]
\mp g_6^{} (2 M_N^2 M_V) (2 E_\Delta \mp M_\Delta)
\biggr\},
\cr
D_{1/2}^{(5/2\pm)} =& \pm \frac{1}{8\sqrt{10}}
\frac{\sqrt{E_\Delta \mp M_\Delta}}{M_N^3 \sqrt{q_V M_N}}
\biggl\{
q_V^2 \left[
g_1^{} (M_R \pm M_\Delta)
\mp g_2^{} \frac{M_R^2 -M_\Delta^2 - M_V^2}{4M_N}
\pm g_3^{} \frac{M_V^2}{2M_N}
\right]
\cr & \hspace{8.0em}
- (E_\Delta \pm M_\Delta)
\left[ g_4^{} \frac{M_R^2 - M_\Delta^2 - M_V^2}{2} - g_5^{} M_V^2 \right]
+ g_6 (2 M_N^2) (2 E_\Delta \pm M_\Delta -M_R)
\biggr\},
\nonumber
\end{align}
%EQUATION<<<
%EQUATION>>>
\begin{align}
A_{3/2}^{(5/2\pm)} =& \frac{1}{16\sqrt{15}}
\frac{\sqrt{E_\Delta \mp M_\Delta}}{M_N^3 \sqrt{q_V M_N}}
\biggl\{
(E_\Delta \pm M_\Delta)
\left[ g_4^{} (M_R^2 - M_\Delta^2 - M_V^2) - g_5^{} (2 M_V^2) \right]
\cr & \hspace{9.5em}
\mp g_6^{}
\frac{4 M_N^2}{M_\Delta}
[(2E_\Delta \pm M_\Delta)(E_\Delta \pm M_\Delta)-(2E_\Delta \mp M_\Delta)(E_\Delta-M_R)]
\biggr\},
\cr
S_{3/2}^{(5/2\pm)} =& -\frac{1}{8\sqrt{10}}
\frac{\sqrt{E_\Delta \mp M_\Delta} M_V}{M_N^3 \sqrt{q_V M_N}}
\left\{
(E_\Delta \pm M_\Delta)
\left[ g_4^{} M_\Delta \mp g_5^{} (M_R \mp M_\Delta) \right] \pm g_6^{} (4 M_N^2)
\right\},
\cr
D_{5/2}^{(5/2\pm)} =& \mp \frac{1}{4}
\frac{\sqrt{E_\Delta \mp M_\Delta}}{M_N \sqrt{q_V M_N}}
g_6^{} (M_R \pm M_\Delta),
%EQUATION<<<
\end{align}
\end{subequations}
where, except for the $g_6$ term, the following relations are satisfied
%EQUATION>>>
\begin{align}
&A_{1/2}^{(5/2\pm)} = \frac{\sqrt{3}}{\sqrt{20}} \frac{q_V}{M_N} A_{1/2}^{(3/2\mp)},\,\,\,
 S_{1/2}^{(5/2\pm)} = \frac{\sqrt{3}}{\sqrt{20}} \frac{q_V}{M_N} S_{1/2}^{(3/2\mp)},\,\,\,
 D_{1/2}^{(5/2\pm)} = \frac{\sqrt{3}}{\sqrt{20}} \frac{q_V}{M_N} D_{1/2}^{(3/2\mp)},
\cr
&A_{3/2}^{(5/2\pm)} = \frac{1}{\sqrt{10}} \frac{q_V}{M_N} A_{3/2}^{(3/2\mp)},\,\,\,
 S_{3/2}^{(5/2\pm)} = \frac{1}{\sqrt{10}} \frac{q_V}{M_N} S_{3/2}^{(3/2\mp)}.
\end{align}
%EQUATION<<<

\textbullet\ spin-$7/2$
\begin{subequations}
%EQUATION>>>
\begin{align}
&A_{1/2}^{(7/2\pm)} = \frac{1}{\sqrt{7}} \frac{q_V}{M_N} A_{1/2}^{(5/2\mp)},\,\,\,
 S_{1/2}^{(7/2\pm)} = \frac{1}{\sqrt{7}} \frac{q_V}{M_N} S_{1/2}^{(5/2\mp)},\,\,\,
 D_{1/2}^{(7/2\pm)} = \frac{1}{\sqrt{7}} \frac{q_V}{M_N} D_{1/2}^{(5/2\mp)},
\cr
&A_{3/2}^{(7/2\pm)} = \frac{\sqrt{5}}{\sqrt{42}} \frac{q_V}{M_N} A_{3/2}^{(5/2\mp)},\,\,\,
 S_{3/2}^{(7/2\pm)} = \frac{\sqrt{5}}{\sqrt{42}} \frac{q_V}{M_N} S_{3/2}^{(5/2\mp)},\,\,\,
 D_{5/2}^{(7/2\pm)} = \frac{\sqrt{2}}{\sqrt{21}} \frac{q_V}{M_N} D_{5/2}^{(5/2\mp)}.
\end{align}
%EQUATION<<<
\end{subequations}

\textbullet\ spin-$9/2$
\begin{subequations}
%EQUATION>>>
\begin{align}
&A_{1/2}^{(9/2\pm)} = \frac{\sqrt{5}}{6} \frac{q_V}{M_N} A_{1/2}^{(7/2\mp)},\,\,\,
 S_{1/2}^{(9/2\pm)} = \frac{\sqrt{5}}{6} \frac{q_V}{M_N} S_{1/2}^{(7/2\mp)},\,\,\,
 D_{1/2}^{(9/2\pm)} = \frac{\sqrt{5}}{6} \frac{q_V}{M_N} D_{1/2}^{(7/2\mp)},
\cr
&A_{3/2}^{(9/2\pm)} = \frac{1}{\sqrt{8}} \frac{q_V}{M_N} A_{3/2}^{(7/2\mp)},\,\,\,
 S_{3/2}^{(9/2\pm)} = \frac{1}{\sqrt{8}} \frac{q_V}{M_N} S_{3/2}^{(7/2\mp)},\,\,\,
 D_{5/2}^{(9/2\pm)} = \frac{1}{3}        \frac{q_V}{M_N} D_{5/2}^{(7/2\mp)}.
\end{align}
%EQUATION<<<
\end{subequations}

\end{widetext}
%------------------------------>>> widetext

%-------------------------------------------------------------------------------------
\section{Summary} \label{sec:summary}
%-------------------------------------------------------------------------------------

We first reviewed the general expression for the propagators of arbitrary spin fields
for bosons as well as for fermions.
It contains explicit formulae for the propagators of several high-spin fields up to
spin-5 (spin-9/2) for bosons (fermions).
This is required to calculate the partial decay widths and to construct the reaction
amplitudes for the hadron scattering processes.

Then, we constructed effective Lagrangians of arbitrary spin baryon resonances which
couple to the following meson-baryon channels
%EQUATION>>>
\begin{align}
& (1a)\, R \to 0^- + \textstyle\frac12^+,   \cr
& (1b)\, R \to 1^- + \textstyle\frac12^+,   \cr
& (2a)\, R \to 0^- + \textstyle\frac32^+,   \cr
& (2b)\, R \to 1^- + \textstyle\frac32^+.
\end{align}
%EQUATION<<<
The process of (1b) is applied to the photo-decay of resonances ($R$) into the
nucleon ($1/2^+$).
In this case, the number of independent couplings reduces by one due to the absence
of the helicity-0 state.
From the constructed effective Lagrangians, we derived the relation between the
coupling constants of effective Lagrangians and the partial decay widths that can be
predicted by hadron models. 
That is, the explicit formulae for the partial-wave decay amplitudes $G(\ell)$ are
given in terms of the coupling constants of effective Lagrangians for the decays of
(1a) and (2a).
Meanwhile, those for the helicity amplitudes $A_\lambda$ are given in terms of the
coupling constants for the decays of (1b) and (2b).
This allows us to compare the coupling constants to the hadron model predictions not
only in magnitude but in sign as well.

In fact, the signs of coupling constants as well as their magnitudes should be
extracted by analyzing observed physical quantities with sophisticated coupled-channel
models.
Our explicit formulae will help facilitate these efforts because the hadron model
predictions can be used as the starting guess for coupling constants in
coupled-channel approaches by using our formulae.
Therefore, more hadron model predictions or experimental measurements of the
partial-wave decay widths and helicity amplitudes will be very valuable to provide
rich information on the mechanism of baryon resonance excitation and to understand
the baryon resonance spectrum.

%-------------------------------------------------------------------------------------
\acknowledgments
%-------------------------------------------------------------------------------------
%We are grateful to K. Nakayama and T.-S. H. Lee for helpful discussions through the
%topical research program of the Asia Pacific Center for Theoretical Physics.
%The authors are grateful to the late Yongseok Oh who initiated this work.
%The authors would like to thank the late Professor Yongseok Oh without whom this work
%would never be completed.
The authors express their deep gratitude to the late Professor Yongseok Oh, whose
leadership was indispensable in initiating and guiding this work.
We are honored to dedicate this study to his memory, completing it under his
inspiration and guidance.
The work was supported by the Basic Science Research Program through the National
Research Foundation of Korea (NRF) under Grants No. 2021R1A6A1A03043957 (S.-H.\,K.,
M.-K.\,Ch.), No. 2022R1I1A1A01054390 (S.-H.\,K.), No. 2020R1A2C3006177 (M.-K.\,Ch.),
and No. 2023R1A2C1004098 (SY.\,S., S.\,S.).

%\appendix
%-------------------------------------------------------------------------------------
\section*{Appendix:  Explicit expressions of propagators} \label{Appendix}
%-------------------------------------------------------------------------------------

The propagator of a field with an arbitrary spin can be written as
%EQUATION>>>
\begin{align}
S(p) = \frac{1}{p^2 - M^2} 
\Delta_{\alpha_1^{} \cdots \alpha_n^{}}^{\beta_1^{} \cdots \beta_n^{}}(j,p)
\end{align}
%EQUATION<<<
for a boson field of spin $j=n$ and
%EQUATION>>>
\begin{align}
S(p) = \frac{1}{p^2 - M^2} (\slashed{p} + M) 
\Delta_{\alpha_1^{} \cdots \alpha_{n-1}^{}}^{\beta_1^{} \cdots \beta_{n-1}^{}}(j,p)
\end{align}
%EQUATION<<<
for a fermion field of spin $j=n-1/2$.
The explicit form of the projection operator 
$\Delta_{\alpha_1^{} \cdots \alpha_n^{}}^{\beta_1^{} \cdots \beta_n^{}}(j,p)$ can be derived from
Eqs.~(\ref{eq:prop-even})-(\ref{eq:prop-fermion}).
Here, we present the explicit formulae for the projection operators of fields upto
spin-$5$ (for bosons) and spin-$9/2$ (for fermions).

Here, we define $\bar{g}_{\mu\nu}$ and $\bar{\gamma}^\mu$ as
%EQUATION>>>
\begin{align}
\bar{g}_{\mu\nu}^{} =& g_{\mu\nu}^{} - \frac{1}{p^2} p_\mu^{} p_\nu^{},
\cr
\bar{\gamma}^\mu =& \gamma^\nu \bar{g}_\nu^\mu =
\gamma^\mu - \frac{1}{p^2} \slashed{p} p^\mu.
\end{align}
%EQUATION<<<
Therefore, $\bar{g}_{\mu\nu}^{} = \bar{g}_{\nu\mu}^{}$ and we have the following identities
%EQUATION>>>
\begin{align}
\gamma_\alpha^{} \gamma_\beta^{} \bar{g}^{\alpha\beta} =& 3,
\cr
\{ \bar{\gamma}_\alpha^{},  \bar{\gamma}_\beta^{} \} =& 2 \bar{g}_{\alpha\beta}^{}.
\end{align}
%EQUATION<<
The results are given as follows

%------------------------------>>> widetext
\begin{widetext}
\textbullet\ \textbf{\boldmath spin-$1$}
%EQUATION>>>
\begin{align}
\Delta_\alpha^\beta(1,p) = -\bar{g}_\alpha^\beta = 
- \left( g_\alpha^\beta - \frac{1}{p^2} p_\alpha^{} p^{\beta} \right).
\end{align}
%EQUATION<<<

\textbullet\ \textbf{\boldmath spin-$1/2$}
%EQUATION>>>
\begin{align}
\Delta \left(\textstyle\frac12,p\right) =
- \frac13 \gamma^\alpha \gamma_\beta^{} \Delta_\alpha^\beta (1,p) = 1.
\end{align}
%EQUATION<<<

\textbullet\ \textbf{\boldmath spin-$2$}
%EQUATION>>>
\begin{align}
\Delta_{\alpha_1^{} \alpha_2^{}}^{\beta_1^{} \beta_2^{}} (2,p) = \frac12
\left( \bar{g}_{\alpha_1^{}}^{\beta_1^{}} \bar{g}_{\alpha_1^{}}^{\beta_2^{}}
+ \bar{g}_{\alpha_1^{}}^{\beta_1^{}} \bar{g}_{\alpha_2^{}}^{\beta_1^{}}
-\frac23 \bar{g}_{\alpha_1^{}\alpha_2^{}}^{}  \bar{g}^{\beta_1^{}\beta_2^{}}
\right).
\end{align}
%EQUATION<<<

\textbullet\ \textbf{\boldmath spin-$3/2$}
%EQUATION>>>
\begin{align}
\Delta_{\alpha_1^{}}^{\beta_1^{}} \left(\textstyle\frac32,p\right) =&
{-\frac{2}{5}} \gamma^\alpha \gamma_\beta^{} \Delta_{\alpha\alpha_1^{}}^{\beta\beta_1^{}} (2,p) 
= - \left( \bar{g}_{\alpha_1^{}}^{\beta_1^{}} 
- \frac13 \bar{\gamma}_{\alpha_1^{}}^{} \bar{\gamma}^{\beta_1^{}} \right)
\cr
=& -g_{\alpha_1^{}}^{\beta_1^{}} + \frac13 \gamma_{\alpha_1^{}} \gamma^{\beta_1^{}}
% + \frac{1}{3M} \left( \gamma_{\alpha_1^{}} p^{\beta_1^{}} - p_{\alpha_1^{}}
% \gamma^{\beta_1^{}}\right) 
+ \frac{\slashed{p}}{3p^2}\left( \gamma_{\alpha_1} p^{\beta_1} - p_{\alpha_1} \gamma^{\beta_1}
\right)
+ \frac{2}{3p^2} p_{\alpha_1} p^{\beta_1}.
\end{align}
%EQUATION<<<

\textbullet\ \textbf{\boldmath spin-$3$}
%EQUATION>>>
\begin{align}
\Delta_{\alpha_1^{} \alpha_2^{} \alpha_3^{}}^{\beta_1^{} \beta_2^{} \beta_3^{}} (3,p) =& 
- \frac{1}{36} \sum_{P(\alpha), P(\beta)}
\left(
\bar{g}_{\alpha_1^{}}^{\beta_1^{}} \bar{g}_{\alpha_2^{}}^{\beta_2^{}}
\bar{g}_{\alpha_3^{}}^{\beta_3^{}}
- \frac35 \bar{g}_{\alpha_1^{} \alpha_2^{}}^{} \bar{g}^{\beta_1^{}\beta_2^{}}
\bar{g}_{\alpha_3^{}}^{\beta_3^{}}
\right)
\cr
=& -\frac{1}{6} \Big(
  \bar{g}_{\alpha_1}^{\beta_1} \bar{g}_{\alpha_2}^{\beta_2} \bar{g}_{\alpha_3}^{\beta_3}
+ \bar{g}_{\alpha_1}^{\beta_1} \bar{g}_{\alpha_2}^{\beta_3} \bar{g}_{\alpha_3}^{\beta_2}
+ \bar{g}_{\alpha_1}^{\beta_2} \bar{g}_{\alpha_2}^{\beta_1} \bar{g}_{\alpha_3}^{\beta_3}
+ \bar{g}_{\alpha_1}^{\beta_2} \bar{g}_{\alpha_2}^{\beta_3} \bar{g}_{\alpha_3}^{\beta_1}
+ \bar{g}_{\alpha_1}^{\beta_3} \bar{g}_{\alpha_2}^{\beta_1} \bar{g}_{\alpha_3}^{\beta_2}
+ \bar{g}_{\alpha_1}^{\beta_3}\bar{g}_{\alpha_2}^{\beta_2}\bar{g}_{\alpha_3}^{\beta_1} \Big)
\cr
 & + \frac{1}{15} \biggl[
  \bar{g}_{\alpha_1 \alpha_2} \Big(
  \bar{g}^{\beta_1 \beta_2} \bar{g}_{\alpha_3}^{\beta_3}
+ \bar{g}^{\beta_1 \beta_3} \bar{g}_{\alpha_3}^{\beta_2}
+ \bar{g}^{\beta_2 \beta_3}\bar{g}_{\alpha_3}^{\beta_1} \Big)
+ \bar{g}_{\alpha_1 \alpha_3} \Big(
  \bar{g}^{\beta_1 \beta_2}  \bar{g}_{\alpha_2}^{\beta_3}
+ \bar{g}^{\beta_1 \beta_3} \bar{g}_{\alpha_2}^{\beta_2}
+\bar{g}^{\beta_2 \beta_3}\bar{g}_{\alpha_2}^{\beta_1} \Big)
\cr
&\quad +
  \bar{g}_{\alpha_2 \alpha_3} \Big(
  \bar{g}^{\beta_1 \beta_2} \bar{g}_{\alpha_1}^{\beta_3}
+ \bar{g}^{\beta_1 \beta_3}\bar{g}_{\alpha_1}^{\beta_2}
+ \bar{g}^{\beta_2 \beta_3}\bar{g}_{\alpha_1}^{\beta_1} \Big)
\biggr].
\end{align}
%EQUATION<<<

\textbullet\ \textbf{\boldmath spin-$5/2$}
%EQUATION>>>
\begin{align}
\Delta_{\alpha_1^{} \alpha_2^{}}^{\beta_1^{} \beta_2^{}} \left(\textstyle\frac52,p\right) =&
-
\frac37 \gamma^\alpha \gamma_\beta^{} 
\Delta_{\alpha\alpha_1^{} \alpha_2^{}}^{\beta\beta_1^{} \beta_2^{}} (3,p) 
\cr
=& \frac12 \left( \bar{g}_{\alpha_1^{}}^{\beta_1^{}} 
\bar{g}_{\alpha_2^{}}^{\beta_2^{}} + \bar{g}_{\alpha_1^{}}^{\beta_2^{}} 
\bar{g}_{\alpha_2^{}}^{\beta_1^{}} \right)
- \frac15 \bar{g}_{\alpha_1^{} \alpha_2^{}}^{} \bar{g}^{\beta_1^{} \beta_2^{}}
- \frac{1}{10} \left( 
\bar{\gamma}_{\alpha_1^{}}^{} \bar{\gamma}^{\beta_1^{}}
\bar{g}_{\alpha_2^{}}^{\beta_2^{}}
+ \bar{\gamma}_{\alpha_1^{}}^{} \bar{\gamma}^{\beta_2^{}}
\bar{g}_{\alpha_2^{}}^{\beta_1^{}}
+ \bar{\gamma}_{\alpha_2^{}}^{} \bar{\gamma}^{\beta_1^{}}
\bar{g}_{\alpha_1^{}}^{\beta_2^{}}
+ \bar{\gamma}_{\alpha_2^{}}^{} \bar{\gamma}^{\beta_2^{}}
\bar{g}_{\alpha_1^{}}^{\beta_1^{}} \right).\cr
\end{align}
%EQUATION<<<

\textbullet\ \textbf{\boldmath spin-$4$}
%EQUATION>>>
\begin{align}
\Delta_{\alpha_1^{} \alpha_2^{} \alpha_3^{} \alpha_4^{}}^{\beta_1^{} \beta_2^{} \beta_3^{}\beta_4^{}}
(4,p) =
\frac{1}{576} \sum_{P(\alpha), P(\beta)}
\left(
\bar{g}_{\alpha_1^{}}^{\beta_1^{}} \bar{g}_{\alpha_2^{}}^{\beta_2^{}} 
\bar{g}_{\alpha_3^{}}^{\beta_3^{}} \bar{g}_{\alpha_4^{}}^{\beta_4^{}}
- \frac67 \bar{g}_{\alpha_1^{} \alpha_2^{}}^{} \bar{g}^{\beta_1^{}\beta_2^{}}
\bar{g}_{\alpha_3^{}}^{\beta_3^{}} \bar{g}_{\alpha_4^{}}^{\beta_4^{}}
+ \frac{3}{35} 
\bar{g}_{\alpha_1^{} \alpha_2^{}}^{} \bar{g}_{\alpha_3^{} \alpha_4^{}}^{}
\bar{g}^{\beta_1^{}\beta_2^{}} \bar{g}^{\beta_3^{}\beta_4^{}}
\right).
\end{align}
%EQUATION<<<

\textbullet\ \textbf{\boldmath spin-$7/2$}
%EQUATION>>>
\begin{align}
\Delta_{\alpha_1^{}\alpha_2^{}\alpha_3^{}}^{\beta_1^{}\beta_2^{}\beta_3^{}}
\left({\textstyle\frac72},p\right)
=& - \frac{4}{9} \gamma^\alpha \gamma_\beta^{}
\Delta_{\alpha\alpha_1^{}\alpha_2^{}\alpha_3^{}}^{\beta\beta_1^{}\beta_2^{}\beta_3^{}} (4,p)
\cr
=& - \frac{1}{36} \sum_{P(\alpha), P(\beta)}
\left(
\bar{g}_{\alpha_1^{}}^{\beta_1^{}} \bar{g}_{\alpha_2^{}}^{\beta_2^{}}
\bar{g}_{\alpha_3^{}}^{\beta_3^{}}
-\frac37
\bar{g}_{\alpha_1^{}}^{\beta_1^{}} \bar{g}_{\alpha_2^{}\alpha_3^{}}
\bar{g}^{\beta_2^{}\beta_3^{}}
-\frac37
\bar{\gamma}_{\alpha_1^{}}^{} \bar{\gamma}^{\beta_1^{}}
\bar{g}_{\alpha_2^{}}^{\beta_2^{}} \bar{g}_{\alpha_3^{}}^{\beta_3^{}}
+ \frac{3}{35}
\bar{\gamma}_{\alpha_1^{}} \bar{\gamma}^{\beta_1^{}}
\bar{g}_{\alpha_2^{}\alpha_3^{}} \bar{g}^{\beta_2^{}\beta_3^{}} 
\right).
\end{align}
%EQUATION>>>

\textbullet\ \textbf{\boldmath spin-$5$}
%EQUATION>>>
\begin{align}
& \Delta_{\alpha_1^{} \alpha_2^{} \alpha_3^{} \alpha_4^{} \alpha_5^{}}^{\beta_1^{} 
\beta_2^{} \beta_3^{}\beta_4^{} \beta_5^{}} (5,p)
\cr &
= - \left( \frac{1}{120} \right)^2 \sum_{P(\alpha),P(\beta)}
\left(
\bar{g}_{\alpha_1^{}}^{\beta_1^{}} \bar{g}_{\alpha_2^{}}^{\beta_2^{}}
\bar{g}_{\alpha_3^{}}^{\beta_3^{}} \bar{g}_{\alpha_4^{}}^{\beta_4^{}}
\bar{g}_{\alpha_5^{}}^{\beta_5^{}}
- \frac{10}{9}
\bar{g}_{\alpha_1^{}\alpha_2^{}}^{} \bar{g}^{\beta_1^{}\beta_2^{}}
\bar{g}_{\alpha_3^{}}^{\beta_3^{}} \bar{g}_{\alpha_4^{}}^{\beta_4^{}}
\theta_{\alpha_5^{}}^{\beta_5^{}}
+ \frac{5}{21}
\bar{g}_{\alpha_1^{}\alpha_2^{}}^{} \bar{g}_{\alpha_3^{}\alpha_4^{}}^{}
\bar{g}^{\beta_1^{}\beta_2^{}} \bar{g}^{\beta_3^{}\beta_4^{}}
\bar{g}_{\alpha_5^{}}^{\beta_5^{}}
\right).
\end{align}
%EQUATION<<<

\textbullet\ \textbf{\boldmath spin-$9/2$}
%EQUATION>>>
\begin{align}
\Delta_{\alpha_1^{} \alpha_2^{} \alpha_3^{} \alpha_4^{} }^{\beta_1^{} \beta_2^{} \beta_3^{}\beta_4^{}}
\left({\textstyle\frac92},p\right)
=& -\frac{5}{11} \gamma^\alpha \gamma_\beta
\Delta_{\alpha\alpha_1^{}\alpha_2^{}\alpha_3^{}\alpha_4^{}}^{\beta\beta_1^{}\beta_2^{}\beta_3^{}\beta_4^{}}
(5,p)
\cr
=& \frac{1}{6336} \sum_{P(\alpha),P(\beta)}
\biggl[
11 \bar{g}_{\alpha_1^{}}^{\beta_1^{}} \bar{g}_{\alpha_2^{}}^{\beta_2^{}}
   \bar{g}_{\alpha_3^{}}^{\beta_3^{}} \bar{g}_{\alpha_4^{}}^{\beta_4^{}}
- \frac{14}{3}
\bar{g}_{\alpha_1^{}\alpha_2^{}}^{} \bar{g}^{\beta_1^{}\beta_2^{}}
\bar{g}_{\alpha_3^{}}^{\beta_3^{}} \bar{g}_{\alpha_4^{}}^{\beta_4^{}}
+ \frac{1}{7}
\bar{g}_{\alpha_1^{}\alpha_2^{}}^{} \bar{g}_{\alpha_3^{}\alpha_4^{}}^{}
\bar{g}^{\beta_1^{}\beta_2^{}} \bar{g}^{\beta_3^{}\beta_4^{}}
\cr
& + \frac{44}{63} \bar{\gamma}_{\alpha_1^{}} \bar{\gamma}^{\beta_1^{}}
\bar{g}_{\alpha_2^{}}^{\beta_2^{}}
( 3 \bar{g}_{\alpha_3 \alpha_4} \bar{g}^{\beta_3 \beta_4} -
7 \bar{g}_{\alpha_3}^{\beta_3} \bar{g}_{\alpha_4}^{\beta_4} )
\cr
& + \frac{4}{21} ( \bar{\gamma}_{\alpha_1} \bar{\gamma}_{\alpha_2} \bar{g}^{\beta_1 \beta_2} +
\bar{\gamma}^{\beta_1} \bar{\gamma}^{\beta_2} \bar{g}_{\alpha_1 \alpha_2} )
(\bar{g}_{\alpha_3 \alpha_4} \bar{g}^{\beta_3 \beta_4} -
 7 \bar{g}_{\alpha_3}^{\beta_1} \bar{g}_{\alpha_4}^{\beta_4} )
\biggr].
\end{align}
%EQUATION<<<
%------------------------------>>> widetext
\end{widetext}
Here the sum is over all possible permutations of $\alpha_i^{}$'s and $\beta_i^{}$'s.
The propagators for higher-spin fields can be obtained in a similar way.

%-------------------------------------------------------------------------------------

%-------------------------------------------------------------------------------------

%\bibliographystyle{/Users/yoh/Dropbox/Document/Research/tex/macros/bibtex/h-physrev4}
%\bibliography{/Users/yoh/Dropbox/Document/Research/biblio/biball}

%-------------------------------------------------------------------------------------
\end{document}